\documentclass[fleqn,usenatbib]{mnras}

\usepackage[utf8]{inputenc}
\usepackage{amsmath,amssymb}
\usepackage{amsfonts}
\usepackage{graphicx}
\usepackage{xcolor}
\usepackage{multirow}
\usepackage{verbatim}
\usepackage[font=footnotesize, labelfont=bf]{caption}

\newcommand{\boldmatrix}[1]{\boldsymbol{\mathsf{#1}}}
\newcommand{\EXP}{{\sc{exp}}}

\pubyear{2023}

\title[Near-equilibrium disc-halo dynamics]{Dynamical Data Mining Captures Disc-Halo Couplings that Structure Galaxies}
\author[A. Johnson et al.]{Alexander Johnson$^{1}$\thanks{E-mail:aj3002@columbia.edu},
Michael S. Petersen$^{2}$,
Kathryn V. Johnston$^{1,3}$,
Martin D. Weinberg$^{4}$\\
$^{1}$Department of Astronomy, Columbia University, 550 West 120th Street, New York, NY 10027, USA\\
$^{2}$Institute for Astronomy, University of Edinburgh, Royal Observatory, Blackford Hill, Edinburgh EH9 3HJ, UK\\
$^{3}$Center for Computational Astrophysics, Flatiron Institute, 162 5th Av., New York City, NY 10010, USA\\
$^{4}$Department of Astronomy, University of Massachusetts, Amherst MA 01003-9305, USA\\}

\begin{document}

\maketitle

\begin{abstract}
Studying coupling between different galactic components is a challenging problem in galactic dynamics.
Using basis function expansions (BFEs) and multichannel singular spectrum analysis (mSSA) as a means of dynamical data mining, we discover evidence for two multi-component disc-halo dipole modes in a Milky-Way-like simulated galaxy.
One of the modes grows throughout the simulation, while the other decays throughout the simulation.
The multi-component disc-halo modes are driven primarily by the halo, and have implications for the structural evolution of galaxies, including observations of lopsidedness and other non-axisymmetric structure.
In our simulation, the modes create surface density features up to 10 per cent relative to the equilibrium model stellar disc.
While the simulated galaxy was constructed to be in equilibrium, BFE+mSSA also uncovered evidence of persistent periodic signals incited by aphysical initial conditions disequilibrium, including rings and weak two-armed spirals, both at the 1 per cent level.
The method is sensitive to distinct evolutionary features at and even below the 1 per cent level of surface density variation.
The use of mSSA produced clean signals for both modes and disequilibrium, efficiently removing variance owing to estimator noise from the input BFE time series.
The discovery of multi-component halo-disc modes is strong motivation for application of BFE+mSSA to the rich zoo of dynamics of multi-component interacting galaxies.
\vspace*{1cm}
\end{abstract}

\section{Introduction}

The structures of galaxies are manifestations of how the laws that govern dynamics combine with the nature of matter. 
Understanding galaxies strengthens our understanding of fundamental physics. 
There are tremendous opportunities to deepen that understanding: a rich legacy of analytic descriptions of galactic dynamics; community investment in high resolution simulations; large scale, high dimensional surveys of billions of stars and galaxies; and the emergence of the vital field of data science to robustly mine and characterise both simulated and real data sets.

Yet recent years have revealed the limits to our conception of our home galaxy, long thought to be a quiet backwater in the Universe. 
Maps of the positions and motions of billions of stars from the {\it Gaia} satellite \citep{Gaia..mission..2016,Gaia..DR2disc..2018, Gaia...DR3...2022}
have revealed a Milky Way in disarray, with abundant signatures of action and reaction - past and ongoing \citep[e.g.][]{Antoja...gaia...2018,Trick..disc..2019,Friske..Schonrich..2019,Helmi..review..2020}. 
These represent significant departures from the descriptions of equilibrium and mild perturbations on which the field of Galactic Dynamics has been built \citep{Binney..Tremaine..2008}. 
Simulations are capable of capturing such complexities but robustly linking the features to theoretical descriptions and identifying their physical origins remains challenging.

Recent work by \cite{Weinberg...mSSA...2021} suggest one approach to this challenge centred around two mathematical tools: Basis Function Expansions (BFE) and Multi-Channel Singular Spectrum Analysis (mSSA). 
{\bf BFE} represent a distribution as a linear combination of basis functions, with half a century of application to galactic dynamics \citep[e.g.][]{cluttonbrock72,cluttonbrock73,kalnajs76,polyachenko81,weinberg89,Weinberg...exp...1999,Petersen...exp...2022}. 
When representing a simulation with a fixed set of basis functions, one obtains time series of coefficients that encode the dynamics in a compressed representation. 
\textbf{mSSA} is a method for identifying temporal correlations. 
Together, one obtains a powerful analysis tool for studying galaxy simulations. The method does not require prior information and thus can be considered a form of unsupervised learning. 
Applying mSSA to BFE time-series, \cite{Weinberg...mSSA...2021} analysed barred-galaxy simulations. 
They found that BFE+mSSA could autonomously extract the dominant space and time correlated features and disentangle different phase of bar formation and evolution recovered through more traditional analysis \citep{Petersen...bar...2021}. 

In this paper, we build on the success of \cite{Weinberg...mSSA...2021} in characterising the evolution of a known feature and explore the use of BFE+mSSA as a dynamical discovery tool. We do so through the analysis of a model galaxy comprised of a stellar disc, stellar bulge, and dark matter halo that is designed to be in equilibrium and hence featureless (described in Section~\ref{sec:methods}). 
Studying such a galaxy serves as a `control' sample for future work with more feature-rich discs, with features from in situ (i.e. spiral arms) or ex situ (i.e. minor mergers) sources. 
With a control model, we want to answer the following questions about BFE+mSSA as a dynamical data mining tool:\\
1) Can BFE+mSSA {\it separate} distinct features that overlap in time and are not distinct by eye (real astrophysical signals, phase mixing, and $N$-body noise)?\\
2) Can BFE+mSSA {\it connect} features within or across components by identifying their shared spatial and temporal structure? \\ 
The answer, as we shall see, is yes to both questions. BFE+mSSA isolates features and allows them to be interpreted independently, while also isolating interactions between components independent of the presence of other interactions.

While analysing the disc in the present study, it became clear that the model was not the perfect featureless system we intended. 
By applying BFE+mSSA to the disc, and then the combination of disc+halo, we identify two dynamical causes of features: phase-mixing from initial conditions, and interactions between the disc and halo.
We identify multiple distinct dynamical signals in each, and examine the dynamical signals in detail (Section~\ref{sec:EvolutionOfNearEquilibriumGalaxy}).
We find that the signals are likely to be generic features in disc+halo systems, and can have real impact on galaxies in the real Universe.

This study is a key step in understanding and exploring the strengths and limitations of BFE+mSSA in multi-component systems (see Section~\ref{sec:LookingAhead}). 
In partnership, BFE+mSSA has great potential beyond simulations analysis. 
Much of analytic linear theory is also built on BFEs. 
Moreover BFEs may be used to described observational data sets. 
Hence BFEs provide a common dynamical language to quantitatively connect theory, simulations, observations and data science while providing rigorous physical interpretations of dynamical processes.
We conclude in Section~\ref{sec:Conclusions} with a discussion of how our results impact galaxy evolution more generally, and how BFE+mSSA fits in a larger program of dynamical data mining.

\begin{figure}
\centering
\includegraphics[width=1.\linewidth]{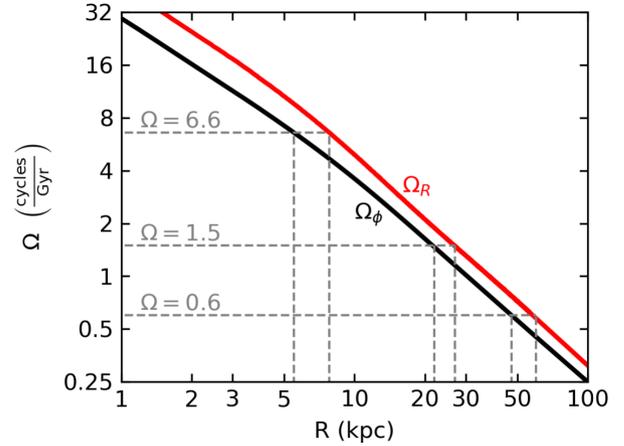}
\caption{Circular (black) and radial (red) frequency curves as a function of radius for the $T=0$ equilibrium model. Both frequencies are computed using the epicyclic approximation, in the plane of the disc ($z=0$). Three frequency values have been marked to guide the eye ($\Omega=0.6$, $\Omega=1.5$, and $\Omega=6.6$ cycles/Gyr), corresponding to spatial scales near the peak disc circular velocity ($2.2R_d=7.7$) and multiples of the halo scale length ($a=52$ kpc).}
\label{fig:Fig1}
\end{figure}

\section{Methods}
\label{sec:methods}

We first review the rationale and overarching goals for BFE+mSSA analysis in dynamical systems in Section~\ref{subsec:RationaleForBFEMSSA}, and then describe the construction of a model isolated disc+bulge+halo galaxy in Section~\ref{subsec:ModelGalaxy}.
Two appendices provide specifics of the expansions used in our analysis (Appendix~\ref{sec:bfedetails}) and an overview of mSSA (Appendix~\ref{sec:mSSA}).

\subsection{Rationale for BFE+mSSA analysis}
\label{subsec:RationaleForBFEMSSA}

All self-gravitating stellar systems, like ionised plasma, have a spectrum of both {\it continuous} and {\it point} modes \citep{Krall.Trivelpiece:73,Ichimaru:73,Ikeuchi.etal:74}.
Here, we define a \emph{mode} to be a superposition of oscillations that lead to a self-similarly growing or damping response to a perturbation\footnote{Mathematically, we are referring to the set of solutions to the collisionless Boltzmann equation for at a specific complex frequency. These are the solutions to the response operator that generalise eigenfunctions in a finite vector space. In plasma physics, these solutions are usually call `modes' although there is some disagreement.}.

{\bf Continuous modes} are excited by perturbations with a continuous range of frequencies, for example a single encounter with a satellite.
Other sources of disequilibrium, whether physical or aphysical, also drive continuous response. This continuous response appears as phase mixing in galaxies.
These modes are also transient: since the response is not dominated by a single frequency the mode quickly looses coherence and therefore is not self-sustaining.
We expect that mSSA will efficiently detect a plethora of signals owing to continuous modes, of varying strength. 
These signals will appear with relatively broad frequency support.
As the modes are transient, few theoretical approaches exist capable of predicting the existence or evolution of these modes, making BFE+mSSA an efficient tool to study them.

{\bf Point modes} are excited by specific frequencies. 
They have model-dependent self-similar shapes and well defined frequencies and can therefore be reinforced by their own gravity.  
The point modes are damped (growing) for stable (unstable) systems.
The most commonly known point mode is the Jeans' instability in a homogeneous sea of stars \citep[e.g.][]{Binney..Tremaine..2008}.
Fluctuations from environmental disturbances such as satellite encounters or Poisson noise from $N$-body distributions may excite these weakly self-gravitating features. 
We expect that some of the results recovered by mSSA will be the phase space manifestation of these modes, appearing as distinct frequency peaks.
Calculations for unstable evolutionary modes in galactic discs have found evidence for point modes supported in various analytic geometries \citep[e.g.][]{Fouvry...discs...2015,DeRijcke...discs...2019}. 
While we do not have explicit theoretical results for damped modes at many azimuthal orders in discs, $N$-body simulations seem to suggest that the amplitude is largest at \(m=2\) and decreases for \(m>2\).
Crucially for the problem at hand (a disc+halo system), we have no analytic predictions for the modal spectra, owing to the complexity of approaching such a problem analytically.
BFE+mSSA gives us a means to detect these modes amongst a sea of other signals.

\begin{figure}
  \centering
  \includegraphics[width=1.\linewidth]{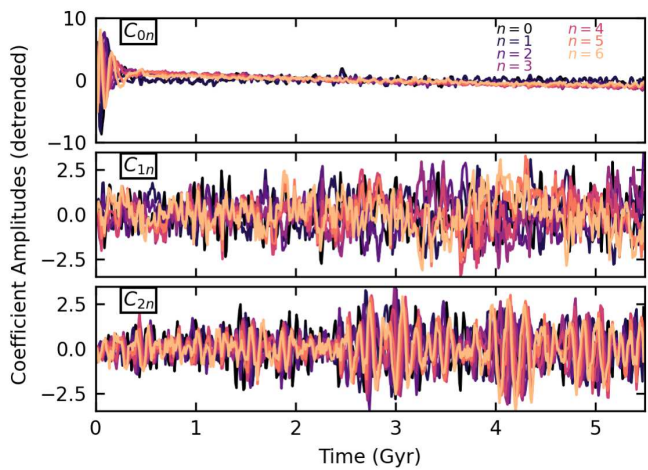}
  \caption{Disc coefficients over time for the first three harmonic orders ($m=0,1,2$) and all corresponding radial orders ($n\in[0,6]$). The coefficients have been detrended by subtracting the mean and dividing out the variance. The coefficient series are dominated by apparent noise, though some trends may be discerned: a steady decrease in some $m=0$ coefficients (upper panel), elevated amplitude towards the end of the simulation in $m=1$, and some periodicity in $m=2$. The origin of these features is difficult to interpret owing to the coefficient series' noisy appearance across multiple basis functions. Any spatial features encoded in the basis are all but impossible to determine.}
  \label{fig:FigA}
\end{figure}

\subsection{Model Galaxy}
\label{subsec:ModelGalaxy}

\subsubsection{Simulation Overview}

We design an isolated model Milky-Way-like galaxy for our study of the compressive power\footnote{Here, `compression' refers to the amount of information one needs to store. A straightforward metric is the total computer disk space. We provide specifics to our simulation, but the scale of compression should be similar in other simulations.} of BFE and the dynamical information one can extract with mSSA. 
We draw the model from components in the merger simulation of \citet{Laporte...SgrSim...2018}: a Hernquist profile dark matter halo with a mass of $10^{12}M_{\odot}$ and a scale length of 52 kpc; an exponential stellar disc with a mass of $6\times 10^{10}M_{\odot}$, a scale length of 3.5 kpc, and a ${\rm sech}^2$ scale height of 0.53 kpc; a Hernquist stellar bulge with a mass of $10^{10}M_{\odot}$ and a scale length of 0.7 kpc. The halo has $40\times10^6$ particles, the disc has $5\times 10^6$ particles, and the bulge has $10^6$ particles.
Unlike \citet{Laporte...SgrSim...2018}, we do not introduce a satellite perturber so that our model galaxy evolves in isolation. 
The initial circular and radial frequency curves in the disc plane are shown in Figure~\ref{fig:Fig1}: as we shall see below, we are able to use these frequencies to inform our mSSA analysis.
We evolve the model with Gadget-4 \citep{Springel...gadget4...2021} for 5.49 Gyr, saving snapshots every 0.01 Gyr, for a total of 549 snapshots. 
The total simulation requires approximately 800 GB of computer disk storage. 

\subsubsection{BFE representation}
\label{sec:bfe}

To compactly describe the simulation, we represent each component in each snapshot with a BFE designed to provide compression and create a continuous representation from the particles. 
Further information regarding the BFEs used may be found in Appendix~\ref{sec:bfedetails}. 
In a BFE, a target distribution is represented as the linear sum of some chosen basis functions, with weighting on each of the basis functions ({\it coefficients}). 
If the basis functions are selected well, the distribution will be described by a small number of functions and corresponding coefficients, $C_\mu$, where $\mu$ is a tag that indexes each basis function. 
The coefficients then are a measure of the importance of each basis function to representing the overall distribution. 
To facilitate representing the distribution with the smallest number of functions, we choose expansions whose lowest-order function resembles the target equilibrium.

For a principally two-dimensional structure, the stellar disc, we use a Fourier-Laguerre expansion\footnote{Another option is presented in \citet{Weinberg...mSSA...2021}: the use of 3d basis functions designed to resemble the exponential disc. In this work, we use the 2d Fourier-Laguerre expansion owing to the straightforward generalisation to the expansion of velocity fields, which will be the subject of future works.}. 
The Fourier-Laguerre basis for expanding disc surface density was introduced in \citet{Weinberg...mSSA...2021}.
Given the exponential weighting of Laguerre polynomials, they serve as a natural radial basis element for exponential discs. 
If the scale lengths are chosen to match, the equilibrium disc is well-represented by the lowest-order Laguerre polynomials. 
The scale length of our Fourier-Laguerre expansion is 3.5 kpc, matching the scale length of the modelled disc. To capture angular structure, we expand in Fourier terms $\cos\phi$ and $\sin\phi$. 
We index the Fourier azimuthal with $m$, and the Laguerre radial terms with $n$, creating $(2m-1)\times n$ total coefficients, each tagged with a unique $(m,n)$, written $C_{mn}$. 
We find that as expected, $C_{00}$ dominates by multiple orders of magnitude as desired. We expand the disc to $m_{\rm max}=6,~n_{\rm max}=6$, making $2\times (m_{\rm max}+1)\times n_{\rm max}=84$ coefficients for the disc.
The choice of maximum radial order is motivated by a desire to probe specific spatial scales. 
The $n=6$ radial Laguerre density function has nodes at 0.9, 3.1, 6.8, 12.1, 19.7, and 30.9 kpc, thus ensuring that the majority of the nodes are within 18 kpc of the disc centre (where 90\% of the particles are located).

% Summary table of features attributable to initial disequilibrium
\begin{table*}
\centering
  \begin{tabular}{|c|c|c|c|c|c|}
                  &     mSSA            &       & DFT peak       & contrast  & SV                \\
 name             &     decomposition   & PCs   & (Gyr$^{-1}$)   & $(R<R_d)$ & fraction          \\
    \hline
    \hline
    \multicolumn{6}{c}{Disequilibrium Signal 1: halo profile readjustment (slow decay)}          \\
    \hline
 Group $m$0-1     & disc $m=0$             & 0,1     & 0.2        & 0.031   & 0.641              \\    
 Group $l$0-1     & halo $l=0$             & 0,1,2,3 & 0.4        & -       & 0.944              \\
 Group $m$0$l$0-1 & disc $m=0$, halo $l=0$ & 0,1,2,3 & 0.2        & 0.054   & 0.832              \\
    \hline
    \hline
    \multicolumn{6}{c}{Disequilibrium Signal 2: phase mixing of disc initial conditions (fast decay)}\\
    \hline
 Group $m$0-2     & disc $m=0$             & 2,3,4,5 & 6.4        & 0.006   & 0.084             \\
 Group $l$0-2     & halo $l=0$             & 4,5     & 6.6        & -       & 0.028             \\
 Group $m$0$l$0-2 & disc $m=0$, halo $l=0$ & 4,5     & 6.6        & 0.007   & 0.037             \\
 Group $m$1-3     & disc $m=1$             & 4,5     & 6.9        & 0.002   & 0.057             \\
 Group $m$2-1     & disc $m=2$             & 0,1     & 6.6        & 0.006   & 0.201             \\
 Group $m$4-1     & disc $m=4$             & 0,1     & 14.2       & 0.004   & 0.086             \\
 Group $m$6-1     & disc $m=6$             & 0,1     & 20.2       & 0.001   & 0.036             \\
 Group $m$2$m$4$m$6-1 & disc $m=2,4,6$     & 0,1     & 6.6        & 0.010   & 0.072             \\
 Group $m$1$l$1-3     & disc $m=1$, halo $l=1$ & 6,7 & 6.9        & 0.002   & 0.040             \\
 Group $m$1$m$2$l$1-2 & disc $m=1,2$, halo $l=1$ & 2,3 & 6.6      & 0.003   & 0.082             \\
\end{tabular}
  \caption{Summary of two different signals identified in our mSSA decompositions as associated with initial disequilibrium. The first signal results from halo disequilibrium, and the appearance in the disc is primarily manifest in the central surface density. The second signal is present in myriad decompositions, but appears to be seeded first by disequilibrium in the disc $m=0$, which then persists in other harmonics. Disc feature strengths are reported in surface density to give a measure of `visual contrast', defined as $\max\left(|\Delta_\Sigma|\right)$ within a disc scale length (see equation~\ref{eq:contrastequation}). Contrasts have an approximate error of 0.001, estimated from grid size adjustments. Owing to simulation sampling rates (0.01 Gyr), the DFT peak is only accurate to 0.1.}
  \label{tab:disequilibrium}
\end{table*}

The dark matter halo\footnote{We also tested bulge expansions, using a similar basis to the dark matter halo. Tests indicated that information contained in the bulge basis was redundant with the dark matter halo: this makes sense for two spherical components. Therefore, we omit the bulge expansion from the analysis in the rest of the paper.} is efficiently described through the empirical orthogonal function basis approach introduced in \citet{Weinberg...exp...1999} and most recently updated in \citet{Petersen...exp...2022}. 
Beginning with the equilibrium distributions, we design a 1d radial model that matches the initial spherically symmetric density profile. 
From this one-dimensional model, we construct an empirical orthogonal function basis whose lowest-order member perfectly matches the input  initial density profile. 
Higher-order terms are generated as eigenfunctions of the Sturm-Liouville equation with the input equilibrium potential-density model and appropriate boundary conditions. 
The three-dimensional structure of the spherical components is described by a spherical harmonic expansion in the angular coordinates. 
Each term in the expansion is represented by three numbers: the spherical harmonic indices $\ell$ and $|m|\le \ell$ and the index of the radial basis function $n$. 
In total, we have $(\ell_{\rm max}+1)^2\times n_{\rm max}$ coefficients per snapshot. 
For the halo, we expand to $\ell_{\rm max}=2,~n_{\rm max}=11$.
The expansions, for the entire simulation, only require approximately 12 MB of storage: a more than $60000\times$ compression, with the benefit of encoding the dynamics.
In practice, we will often consolidate the same-integer positive and negative spherical harmonic $m$ indices when describing the coefficient amplitudes such that a quoted $(\ell,m)$ tag contains both $\pm m$. 
As expected, the $C_{\ell m n}=C_{000}$ term is the largest by multiple orders of magnitude, with $C$ generally decreasing as either $(\ell,m)$ or $n$ increases.

\begin{figure*}
  \centering
  \includegraphics[width=.95\linewidth]{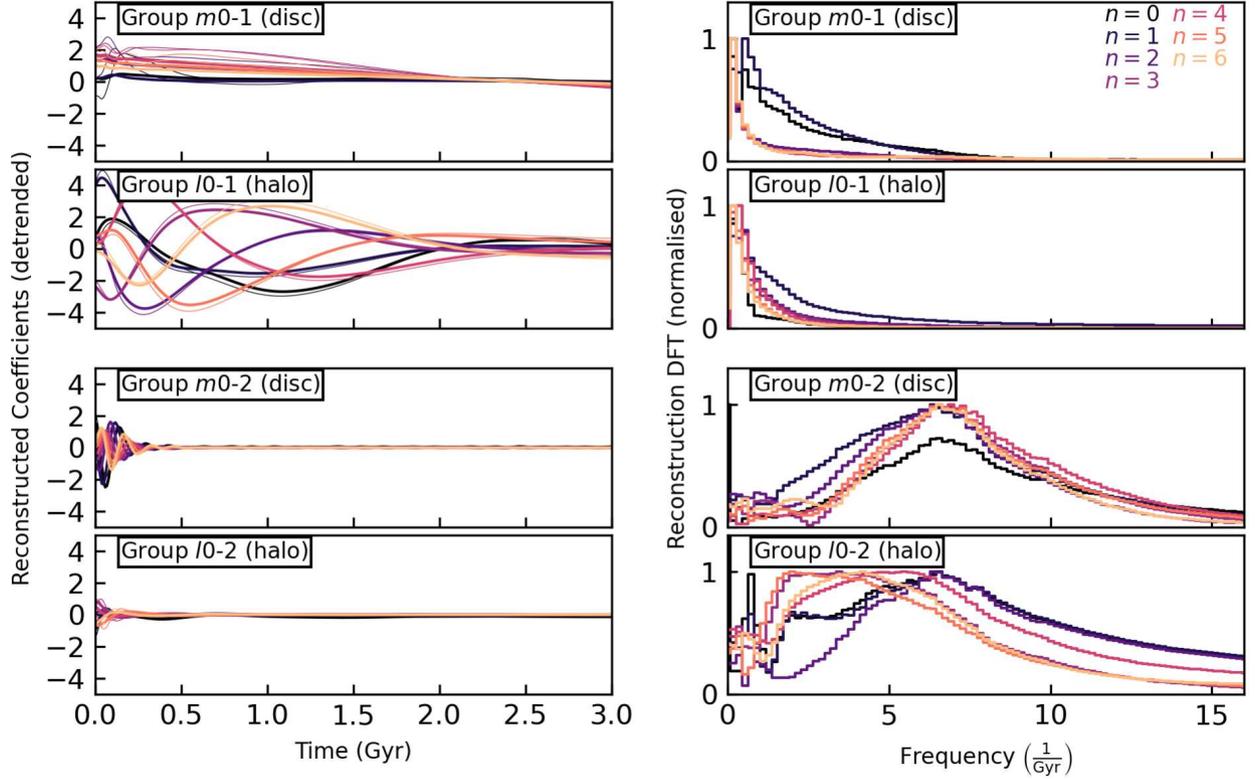}
  \caption{An analysis of two monopole signals resulting from distinct sources of initial disequilibrium. The left panels show the reconstructed coefficient amplitudes over time for each signal (identified as Groups 1 and 2 in both disc-only, halo-only, and disc+halo analyses). The right panels show the power spectra of the reconstructed coefficients for each group. The first signal is a slow rearrangement owing to the halo settling in the presence of the disc, manifest by eye in the disc primarily as a change in the central surface density (cf. Figure~\ref{fig:FigG}). We show the appearance of this signal in the disc and halo as the upper two rows. The second signal is ringing in the disc resulting from the initial velocity disequilibrium of the disc. While the signal decays rapidly in the monopole component, the disequilibrium seeds long-lasting persistent periodic features in other harmonics: see entries under `Disequilibrium Signal 2' in Table~\ref{tab:disequilibrium}. We show the appearance of this signal in the disc and halo as the lower two rows. In each left-hand panel, we show two thicknesses of curves: the thick lines are for the components when analysed separately and the thin lines are for the components when analysed jointly. That the different thicknesses of lines, for the same radial order, are not particularly different, is strong evidence that the features are correlated between the disc and halo.}
  \label{fig:FigH}
\end{figure*}

\begin{figure}
  \centering
  \includegraphics[width=1.\linewidth]{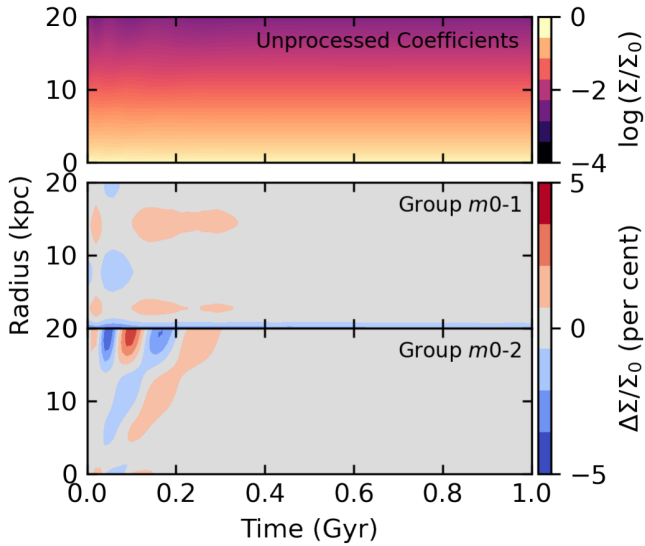}
  \caption{Disc monopole ($m=0$) surface density as a function of radius and time, computed from the full coefficient series (upper panel), showing a largely featureless disc. The surface density has been normalised by the central surface density. The remaining panels show the contribution to the surface density deviations for two groups of $m=0$ principal components, identified as two disequilibrium signals (see Table~\ref{tab:disequilibrium}). The surface density deviations are computed relative to the $m=0,n=0$ background, and are of the order a few per cent (excepting the outer disc, where the low densities mean a variations naturally result in a larger per cent variation).}
  \label{fig:FigG}
\end{figure}

\begin{table*}
\centering
  \begin{tabular}{c|c|c|c|c|c|}
                  & mSSA            &         & DFT peak     & contrast  & singular value \\
 name             & decomposition   & PCs     & (Gyr$^{-1}$) & $(R<R_d)$ & fraction       \\
    \hline
    \hline
    \multicolumn{6}{c}{Point Mode 1: slow growth}\\
    \hline
 Group $m$1-1     & disc $m=1$             & 0,1     & 0.6          & 0.007     & 0.201          \\
 Group $l$1-1     & halo $l=1$             & 0,1,2,3 & 0.4          & -         & 0.272          \\
 Group $m$1$l$1-1 & disc $m=1$, halo $l=1$ & 0,1,2,3 & 0.6          & 0.008     & 0.244          \\
    \hline
    \hline
    \multicolumn{6}{c}{Point Mode 2: slow decay}\\
    \hline
 Group $m$1-2     & disc $m=1$             & 2,3     & 1.7          & 0.003     & 0.064             \\
 Group $l$1-2     & halo $l=1$             & 4,5     & 1.5          & -         & 0.035             \\   
 Group $m$1$l$1-2 & disc $m=1$, halo $l=1$   & 4,5     & 1.5          & 0.003     & 0.048             \\
\end{tabular}
  \caption{The coupled disc+halo dipole modes appearing in different mSSA decompositions. Both modes appear in multiple mSSA decompositions, and that they both appear in disc-only, halo-only, and disc-halo decompositions strongly suggests that they ar both joint modes. In the table, disc harmonics are denoted with $m$, halo harmonics are denoted with $l$. Columns are the same as in Table~\ref{tab:disequilibrium}.}
  \label{tab:realdynamics}
\end{table*}

\begin{figure*}
  \centering
  \includegraphics[width=1.\linewidth]{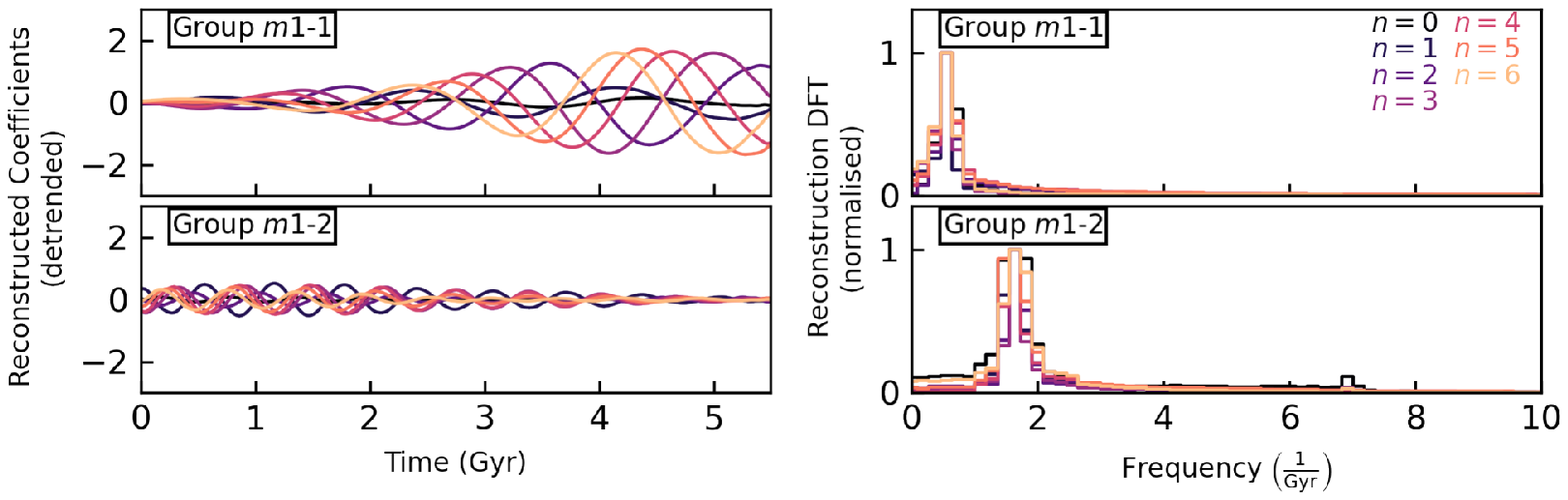}
  \caption{An analysis of two groups obtained from the disc-only $m=1$ mSSA decomposition. Each group corresponds to a distinct point mode, discussed in the text as `Mode 1' and `Mode 2'. The left panels show the reconstructed $m=1$ coefficient amplitudes over time for Groups $m1$-1 and $m1$-2. The right panels show the power spectra of the reconstructed $m=1$ coefficients for each group. Both modes have well-defined slow patterns -- significantly slower than any frequency associated with stars in the disc -- and show evolving behaviour: the first mode is unstable and grows with time, while the second mode is damped and decays with time. The mode summaries are listed in Table~\ref{tab:realdynamics}.}
  \label{fig:FigB}
\end{figure*}

\section{Evolution of a near-equilibrium galaxy}
\label{sec:EvolutionOfNearEquilibriumGalaxy}

Our isolated disc+bulge+halo galaxy was constructed to be in a completely stable equilibrium. 
However, the model is not in equilibrium, for reasons both physical and unphysical.
Figure~\ref{fig:FigA} shows the raw BFE coefficients for the low-order disc harmonics derived from the simulation snapshots.
While it is clear that the coefficient time-series are noisy, inspection by eye suggests that there exists lower frequency coherent signals buried in the higher frequency noise: 
early evolution in $m=0$; modestly elevated power at late times in $m=1$; and a periodic signal in $m=2$. 

To explore dynamical evolution in our simulation, we performed mSSA decompositions of various combinations of BFE coefficients. 
These decompositions revealed clean, persistent features in the individual low-order disc harmonics ($m=0,1,2$), which we concentrate on understanding in this section. 
We also augment the analysis of the low-order disc harmonics with mSSA analysis of halo coefficients, joins of disc and halo coefficients, and higher-order disc harmonics ($m>2$).
These multi-component mSSA analyses prove to be the most fruitful in identifying the causes of different features.
The full results of all our analyses are presented in Appendix~\ref{appendix:allfeatures}.

Section~\ref{subsec:use} describes how the results of the mSSA analysis can be used to group coefficients into separate dynamical features, characterise the properties of these features and come to a physical understanding of their nature.
The following subsections illustrate these ideas by
dividing our own analysis of the disc+bulge+halo simulation into three classifications: initial conditions disequilibrium (Section~\ref{subsec:InitialConditionsDisequilibrium}), secular evolution signals (Section~\ref{subsec:SecularEvolutionSignals}),  and  fluctuations and other uninterpretable features (Section~\ref{subsec:Nullity}).

\begin{figure*}
  \centering
  \includegraphics[width=1.\linewidth]{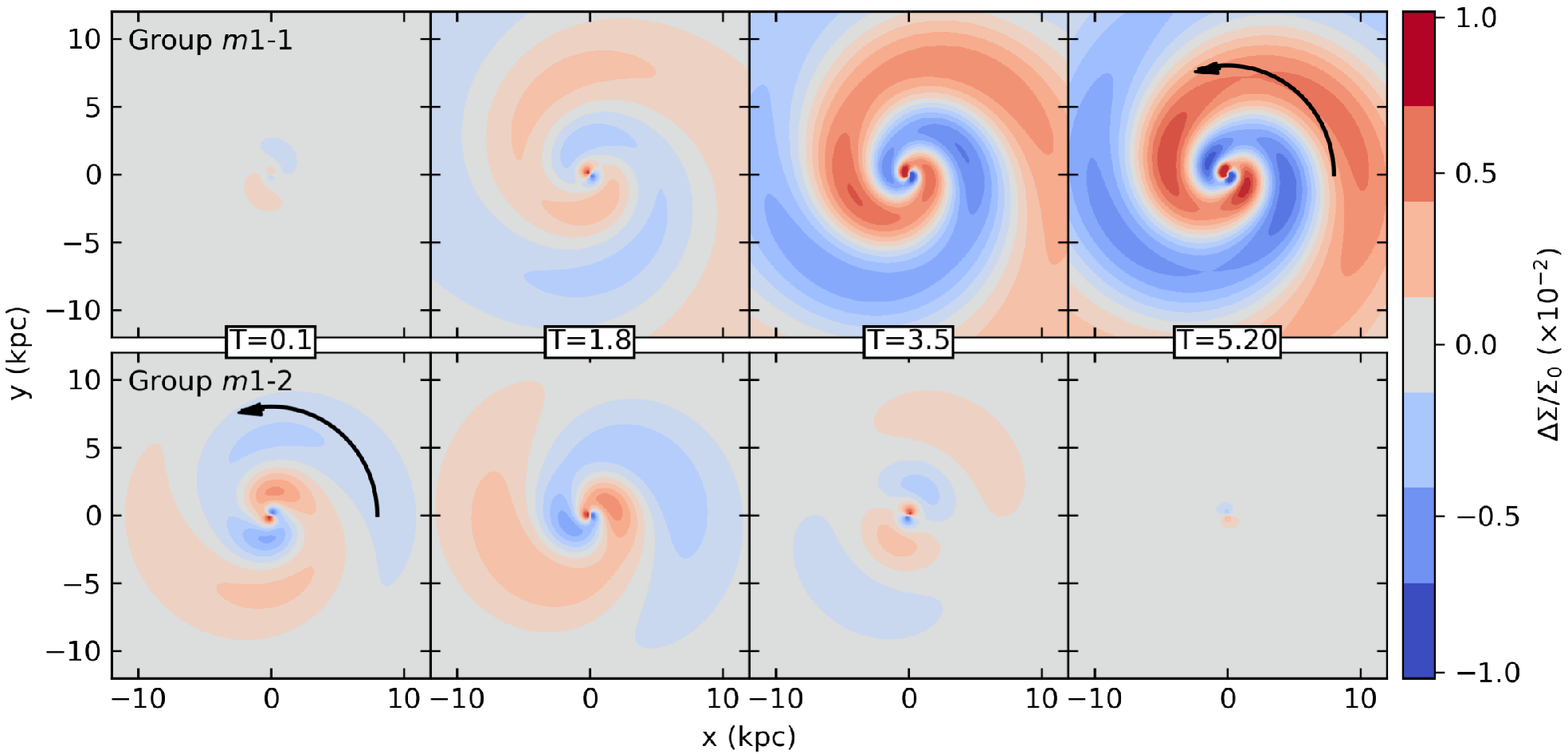}
  \caption{Normalised face-on $(x,y)$ disc surface density deviation determined for two groups in the $m=1$ decomposition. Each group corresponds to a distinct point mode, discussed in the text as `Mode 1' and `Mode 2'. The panels shows a reconstruction of snapshots for either Group $m1$-1 (upper row) or Group $m$1-2 (lower row) in the disc-only $m=1$ decomposition (cf. Figure~\ref{fig:FigB}). Both groups are retrograde with respect to the disc rotation (rotation direction of the pattern is marked with an arrow). The mode shown in the upper panels grows in amplitude over the course of the simulation; the mode shown in the lower panels decays in amplitude over the course of the simulation, evident from the surface density features. Neither pattern strongly winds; both are a largely self-similar evolution, despite being fairly tightly wound.}
  \label{fig:FigC}
\end{figure*}

\begin{figure}
  \centering
  \includegraphics[width=1.\linewidth]{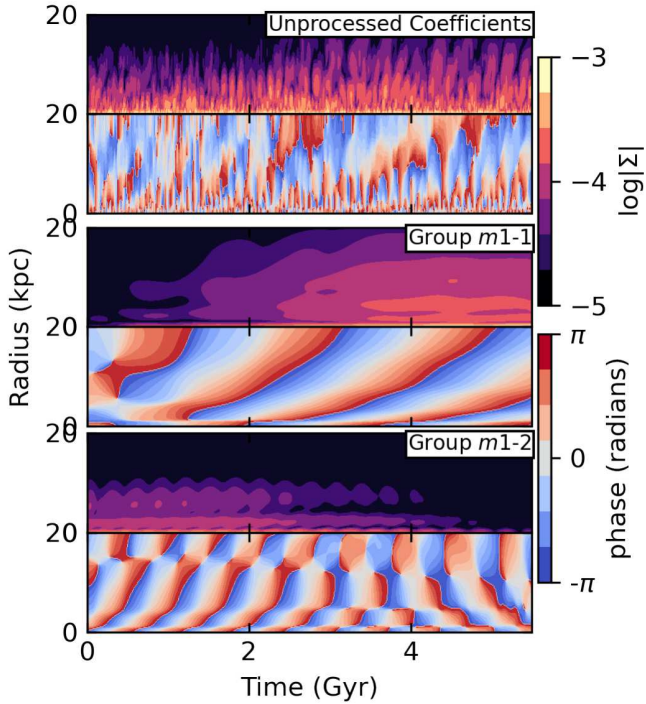}
  \caption{Amplitude and phase as a function of radius and time for the disc-only $m=1$ decomposition for the first two groups identified in the mSSA analysis. Each group corresponds to a distinct point mode. From top to bottom, we show the amplitude and phase for the unprocessed $m=1$ coefficient streams, the reconstructed coefficients of Group $m$1-1, and the reconstructed coefficients of Group $m1$-2. The density is shown as the log of the absolute value of the density. Both groups show coherent phases identifiable in the seemingly random phase information of the unprocessed coefficients. The growing (decaying) nature of Group $m$1-1 (Group $m$1-2) is also evident in the amplitudes.}
  \label{fig:FigD}
\end{figure}

\begin{figure}
  \centering
 \includegraphics[width=1.\linewidth]{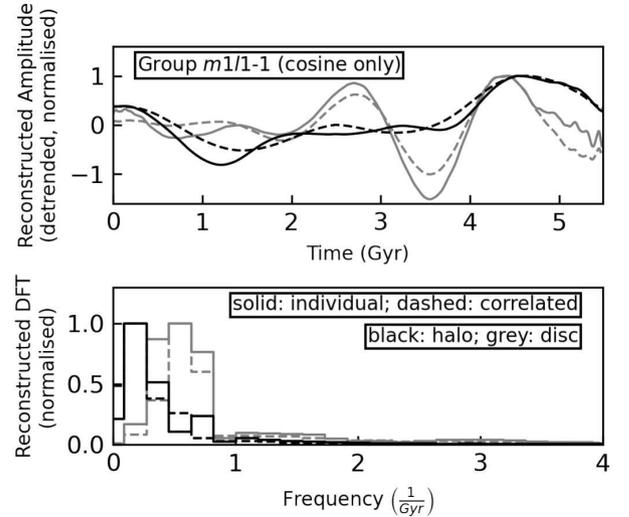}
  \caption{Description of the strongest principal component group for halo and disc decompositions: a growing multi-component point mode. The upper panel shows the detrended and normalised amplitude of the reconstructed cosine component of the $m=1$ (disc; grey curves) or $l=1$ (halo; black curves) $n=0$ coefficient versus time. The solid curves are for mSSA decompositions run on each component alone (Group $m$1-1 and Group $l$1-1). The dashed curves are for the joint halo+disc mSSA decomposition (Group $m$1$l$1-1). The lower panel shows the power spectrum (DFT amplitude vs frequency), for the four series shown in the upper panel. The relative similarity of the curves and power spectra suggests that the patterns are correlated between the disc and halo. The slow growth of the disc amplitude over time relative to the larger halo amplitude at the outset of the simulation suggests that the halo is responsible for driving the mode.}
  \label{fig:FigF}
\end{figure}

\begin{figure*}
  \centering
  \includegraphics[width=1.\linewidth]{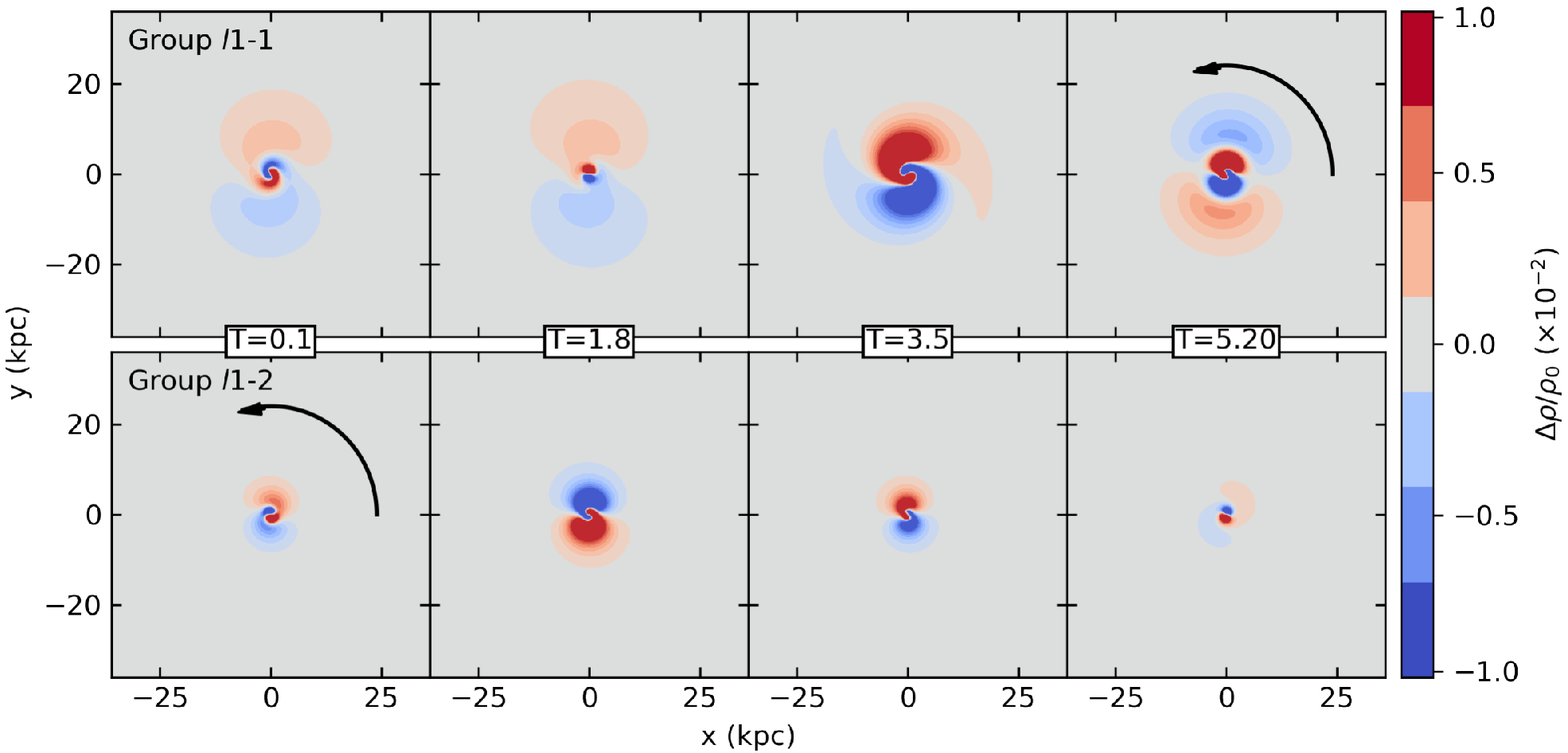}
  \caption{Normalised face-on $(x,y)$ halo $z=0$ plane density deviation reconstruction snapshots for Group $m$1$l$1-1 (upper panels) and Group $m$1$l$1-2 (lower panels) in the halo-and-disc $l=1+m=1$ decomposition. Each group corresponds to a distinct point mode. The patterns extends to large radii in the halo and are retrograde with respect to the disc rotation. The halo reconstructions exhibit significantly less ordered behaviour compared to the disc owing to the three-dimensional nature of the mode, which also tips relative to the $z=0$ plane. However, the bulk properties are similar to the disc (cf. Figure~\ref{fig:FigC}). The mode summaries are listed in Table~\ref{tab:realdynamics}. That the joint decomposition of the halo and disc returns the same groups, with similar behaviour, is strong evidence for the mutual mode nature of the features. The large spatial scale of the modes in the halo, coupled with their relatively early coherence, is suggestive that the modes are induced by the halo.}
  \label{fig:FigE}
\end{figure*}

\subsection{Interpreting the results of the mSSA analysis}
\label{subsec:use}

We use several diagnostics (denoted below in slanted text) to describe the character and understand the nature of the features identified in the mSSA analysis. Each diagnostic has a corresponding section in Appendix~\ref{sec:mSSA} describing the mathematical details.

Applied to BFE multiple series, mSSA identifies temporally correlated signals in the BFE coefficients series as an ensemble. Briefly, mSSA uses the autocorrelation of time lagged matrix of the input series and performs an eigenanalysis to find dominant trends. Each time series is detrended by its mean and variance to intercompare the variations in each coefficient series with. 
These eigenvectors describing these trends are usually called \textit{principal components} (PCs). 
As we always find multiple PCs contribute to a single dynamical feature in our analysis (see `PCs' column in Tables), we will refer to each feature as a `Group' (of PCs), labelling the strongest group (ordered by PC variance) as the first group.
We also denote the particular decomposition by the input coefficient harmonic in the group name.
For example, the strongest group in the $m=0$ disc analysis will be labelled `Group $m$0-1', and the strongest group in the $l=1$ halo analysis will be labelled `Group $l$1-1'.
As PC groups capture trends in basis function coefficients that are correlated over snapshots, PC groups capture how spatial features dynamically evolve. 

Mathematically descriptive (but often difficult to interpret beyond the most significant few), the mSSA decomposition returns {\it singular values} (SVs) as measurements of the contribution of each PC to the total decomposition. 
Larger SVs indicate which PCs represent more of the net change in time of the distribution. 
This property greatly helps the robust identification of features that represent true dynamical evolution. 
PCs which correspond to random fluctuations due to (e.g.) numerical noise are by nature uncorrelated. 
They have very low SV even as they may be the dominant source of variations in the surface density. 
Conversely, PCs which describe evolution in coefficient series that are coherent over time will have high SV even though they may be (orders of magnitude) below the inherent noise. 
We report the singular value fraction\footnote{To compute the relative contribution, we normalise each singular value corresponding to a particular principal value to the sum of all singular values. Then, we can say that some per cent of the signal is represented by the principal component (or group). We will call this the contribution of a principal component (or group), and may be interpreted as a measure of signal robustness.} attributable to a given group in the Tables.

We examine the \textit{coefficient reconstructions} from a group of PCs for physical insight. 
From the coefficient reconstructions, we can also construct \textit{power spectra} from a Discrete Fourier Transform (DFTs) of the reconstructed coefficients from a group of PCs give insight into frequencies (and time scales) that characterise the time evolution of a feature. 
Approximately equal values of dominant frequencies in the power spectra of the coefficient reconstructions between different PCs from mSSA of the same component suggest they are describing different aspects of the same feature and may be grouped together. 
If equal values occur across different components they may be mutually interacting. 
See the `DFT peak' entry in Tables, which reports the frequency value where the DFT is maximised.

We can also calculate {\it contrast} in the disc from the reconstructions\footnote{We do not look at the  contrast in the halo, as this is not straightforwardly measured in real galaxies. Therefore, the contrast columns do not contain entries for halo-only mSSA analyses.}.
Calculating the average of the fractional deviation in surface density within one disc scalelength gives a measure of the `detectability' of a feature (by eye or algorithm). 
See the `contrast' entry in Tables. 
Related, the inferred location in the galaxy is where the dominant frequencies found in the power spectrum match the circular velocity of the unperturbed galaxy can indicate the spatial scales of any interactions taking place. 
Refer to Figure~\ref{fig:Fig1}.

In general, identified features evolve as one of the following types of evolution (noted in Tables): decaying, where a feature peaks at the beginning of the simulation and decays in importance; growing, where the feature grows and then saturates in amplitude with later maximum times therefore having slower growth rates; or consistent with no evolution. 
By comparing the evolution type across different components, one may also infer causality. 
The relative growth or decay may indicate when one component is driving another.

\subsection{Initial Conditions Disequilibrium Uncovered Through Disc $m=0$ Analysis}
\label{subsec:InitialConditionsDisequilibrium}

We start our investigation with perhaps the most striking feature in the raw coefficients apparent in the top panel of Figure~\ref{fig:FigA} which shows the evolution of the $m=0$ (monopole) disc coefficients. 
The figure suggests the simulation suffers from a disequilibrium that is typical in disc-halo initial conditions: outwardly propagating rings in surface density.
This section reports the insights into this apparent evolution afforded by mSSA, starting from its application to the $m=0$ disc coefficients alone (\ref{subsec:m0disc}). 
The properties of the features identified in this preliminary analysis provide a template for further applications of mSSA both to the halo (separately and combined with the disc, see \ref{subsec:HaloDrivenDisequilibrium}) and higher order disc terms (see \ref{subsec:DiscDrivenDisequilibrium}).
Table \ref{tab:disequilibrium} summarises the properties of all these analyses.

\subsubsection{Grouping into Dynamical Features}
\label{subsec:m0disc}

The mSSA analysis of the $m=0$ disc reconstructed coefficients reveals that PCs (0,1) and PCs (2,3,4,5) had distinct power spectra, suggesting natural groupings. 
This also suggested the presence of \textit{two} distinct dynamical features with the signal in Figure~\ref{fig:FigA}. 
The properties of these two groups that are quoted below are summarised in Table~\ref{tab:disequilibrium}, with the rows labelled `Group $m0$-1' and `Group $m0$-2' corresponding to this first mSSA analysis.

Two more figures illustrate our results. 
Figure~\ref{fig:FigG} shows the amplitude (left hand panel) and DFTs of the coefficient reconstructions for Groups $m0$-1 and $m0$-2, revealing their distinct temporal characteristics.
In Figure~\ref{fig:FigH}, we show the $m=0$ surface density amplitude reconstruction as a function  of disc radius (y-axis) and time (x-axis) from the unprocessed coefficients (top panel), as well as the surface density deviations relative to a smooth monopole background, constructed from the two $m=0$ PC groups.

Overall, we find the following characteristics. \\
\underline{Group $m0$-1} represents a dynamical feature that shows weak evolution over the entire simulations with a 
surface density contrast of approximately 3 per cent. The slow decay of Group $m0$-1 produces power at a range of very low frequencies, peaked at 0.2Gyr$^{-1}$.\\
\underline{Group $m0$-2} shows outwardly propagating rings in surface density that start at the beginning of the simulation and disappear after $\approx1$ Gyr, losing speed as they move to larger radii. 
While this is a sub-1 per cent effect within a disc scale length, at larger radii, the surface density deviation is obvious by eye as ringing features. The periodic nature of 
Group $m0$-2 corresponds to a frequency peak at 6.4Gyr$^{-1}$.

{\it We conclude that mSSA has cleanly separated two distinct evolutionary processes operating simultaneously within one harmonic term.} The next two subsections explore the nature of both of these features.

\subsubsection{Group 1: Halo-driven disequilibrium?}
\label{subsec:HaloDrivenDisequilibrium}

The appearance of Group $m0$-1, at low frequency, suggests that its origin may be connected to the halo, where timescales are naturally long. 
Specifically, the frequency 0.2 Gyr$^{-1}$ corresponds to a circular orbit  at $R~\sim~50$~kpc (see Figure \ref{fig:Fig1}). 
This motivated us to apply mSSA to the $l=0$ coefficients representing the halo component in the simulation to explore this connection further. 
We run analyses of both the halo $l=0$ alone and in combination with the disc $m=0$ coefficients. 

The results of the analysis of the halo alone is shown in lower panels of Figure~\ref{fig:FigH} and summarised in the second row of Table~\ref{tab:disequilibrium}.
These demonstrate that the readjustment of the halo component's radial profile is even more significant than the disc radial profile, with a signal amplitude twice as strong as the disc (compare detrended amplitudes in Figure~\ref{fig:FigH}). 
Such halo-driven disequilibrium is also a common feature for numerical realisations of multi-component galaxies as their combined equilibrium properties have been approximated, for example through Jeans modelling or adiabatic contraction corrections. 
Thus the mass distribution of the halo adjusts to full equilibrium in the presence of the disc, and vice versa.  

In Table~\ref{tab:disequilibrium}, a comparison of rows 1 (analysis disc coefficients alone), 2 (halo coefficients alone) and 3 (disc and halo coefficients combined) confirms: (i) all three mSSA analyses have similar temporal structures, corresponding to the dynamical timescales at several tens of kpc in the system; (ii) the joint disc/halo analysis actually identifies the same coherent features in the disc and at greater contrast (0.054 vs 0.031) than the disc analysis alone; (iii) that the driver for the combined evolution is likely the halo given the larger amplitude of its coherent changes relative to random fluctuations for that component.

{\it The above results demonstrate the ability of mSSA to successfully identified the mutual readjustment of the coupled disc-halo system from a mild disequilibrium state. }

\subsubsection{Group 2: Disc-driven disequilibrium}
\label{subsec:DiscDrivenDisequilibrium}

The strength of Group $m$0-2 in the analysis inspired an investigation as to whether this disequilibrium could also seed other features in the simulation. 
Examination of other mSSA decompositions for different coefficient combinations finds many similar-frequency signals (see lower rows of Table~\ref{tab:disequilibrium}). 
Even disc harmonics ($m=2,4,6$) show a persistent signal in the most important PCs (0 and 1) with a pattern speed of $\sim 3.3$ cycles/Gyr that is equal to the half the Group $m0$-2 frequency peak of $6.6$ cycles/Gyr\footnote{The pattern speed of a harmonic is the number of cycles per Gyr divided by the harmonic number. That is, the pattern speed of the disc-only decomposition of Group 2 $m$ harmonic coefficients is $\Omega_{m} = \Omega_{\rm DFT}/m$ cycles/Gyr.}. 
Note that the joint analysis of all even disc harmonics ($m=2,4,6$) returns essentially the same results as the $m=2$ only decomposition.
In the case of harmonic orders $m>2$, this result likely owes to the need for higher order harmonics to fully represent the feature being described.
 
The remaining rows of Table~\ref{tab:disequilibrium} demonstrate that the Group $m0$-2 disc disequilibrium signal is also evident at a lower level (i.e. higher PC numbers, lower contrast in the disc and smaller SV) in both the disc $m=1$ and halo $l=1$ decompositions when comparing frequency structure of the groups. 
While the peak surface density deviation is near the outset of the simulation for $m=0$, in higher harmonic orders the signal does not completely fade over the simulation, with peak measured contrasts coming at later times. {\it Our findings show the utility of mSSA in detecting evolution incited across different harmonic orders.}

\subsubsection{Key insights}

In this section, BFE+mSSA has been used to increase our understanding of a dynamical simulation by: \\
(i) separating distinct evolutionary pathways within a single harmonic; \\
(ii) identifying coupling between multiple components; \\
(iii) detecting features across different harmonics within a single component.

These results emphasise that initial conditions for near equilibrium studies of galaxy evolution need to be dynamically relaxed (or virialised) by evolving in isolation for tens of halo dynamical times (i.e. much longer than than the equivalent timescale in the disc) prior to an studies of interactions in order to truly isolate signatures of the external perturbation. While the perturbation in our study is a numerical artifact, the distinct adjustments to density profiles and couplings within and across components uncovered by BFE+mSSA represent the drivers of the evolution of galaxies seeded by any perturbation. 

\subsection{Secular evolution signals Uncovered Through Disc $m=1$ Analysis} 
\label{subsec:SecularEvolutionSignals}

Examination of the PCs from the mSSA decomposition of the dipole disc harmonic ($m=1$) revealed two groups, with properties summarised in Table \ref{tab:realdynamics} and contributing coefficients and power spectra visualised in the left and right panels of Figure \ref{fig:FigB}. 
Examination of the power spectra show that these features are distinct in nature to the disequilibrium-seeded $m=0$-dominated Groups $m0$-1 and $m0$-2 described in the previous section in that they have clear, well-defined frequencies, rather than a broad spectrum.
This indicates that each of these groups may be a {\it point mode} present in the system.
As discussed in Section~\ref{subsec:RationaleForBFEMSSA}, point modes are a result of the fundamental properties of the underlying phase-space distribution. 
They have single-valued real and imaginary frequencies (hence the descriptive \emph{point}) that describe the periodicity and growth or decay of the features they support. 
These modes drive secular, self-sustained evolution distinct from that of a transient response to an external driver (e.g. the disequilibrium initial conditions in the previous section) that phase mixes away. 
Hence we refer to these groups as `Mode 1' and `Mode 2', and examine their nature in the following subsections. In the disc $m=1$ analysis, these are Groups $m1$-1 and $m1$-2.

\subsubsection{Appearance of modes in the disc}
\label{subsubsec:DiscAppearance}

We augment the information about the two modes summarised in Figure~\ref{fig:FigB} and Table~\ref{tab:realdynamics} with visualisations of their appearance in Figures~\ref{fig:FigC} and \ref{fig:FigD}.
Figure~\ref{fig:FigC} shows selected face-on disc surface density reconstructions to demonstrate that both modes create spiral patterns that are retrograde relative to the rotation of the disc. 
Figure~\ref{fig:FigD} illustrates the radial (y-axis) and time (x-axis) evolution of the surface density (upper panel in each pair) and phase over (lower panel in each pair) for the full time sequence, indicating both the growth/decay and periodicity.
Inspection of these figures and the table provide the full characterisation of the modes.

{\bf Mode 1} groups $m=1$ PCs 0 and 1, reconstructing a slowly rotating, growing mode. Referring to Figure~\ref{fig:Fig1}, the frequency of the signal ($\Omega=0.6$ cycles/Gyr) is located near the scale radius of the halo, well outside the disc\footnote{For $m>0$ harmonics, PC groupings frequently occur in pairs that describe both the amplitude and phase of a feature. In the left hand panels of Figure~\ref{fig:FigB} only the cosine terms in the coefficients are plotted to allow the reader to infer both amplitude and periodicity.}.
Mode 1 grows significantly in amplitude over the simulation, with the peak surface density signal coming near the end of the simulation. 
Computing the contrast in the outer, low-density disc ($r>12$ kpc), the surface density deviation amplitude reaches 10 per cent, detectable as lopsidedness in deep imaging of disc galaxies. 

{\bf Mode 2} groups $m=1$ PCs 2 and 3, reconstructing a slowly rotating, slowly decaying mode. 
The frequency of the signal ($\Omega=1.7$ cycles/Gyr) is located closer to the Galactic centre, but also beyond the bulk of the disc mass.
Mode 2 decays from the outset of the simulation, and is significantly weaker than the first mode, with a peak contrast of order 0.1 per cent within a scale length.

\subsubsection{Connection between the disc and halo}
\label{subsubsection:DiscHaloConnection}

Since the frequencies of the two modes are consistent with halo frequencies we naturally suspect that the halo is supporting the modes. 
To test this, we perform additional mSSA decompositions: first with the $l=1$ halo coefficients alone, and then with the $l=1$ halo coefficients jointly with the $m=1$ disc coefficients\footnote{To find correlated features between the halo and disc we choose halo coefficients that can describe features with meaningful projections into the disc plane. To this end we choose only the $Y_l^m=Y_1^1$ terms of the halo expansion, excluding the $Y_1^0$ term. In addition we use the same number of coefficients from each component to avoid introducing the prior of unequal representation.}. 
The results of the runs are summarised in Table~\ref{tab:realdynamics}. 
We find sets of PCs in the halo-only decompositions corresponding to Modes 1 and 2, which we associate by means of their similar frequencies.
We also find corresponding PCs in the joint disc-halo decomposition.
The joint analysis in particular suggests that the modes are multi-component in nature, owing to the similar properties between all decompositions.

Figure~\ref{fig:FigF} provides an example visualisation for a single radial coefficient ($n=0$) contributing to Mode 1 to verify this interpretation.
Comparing the coefficients reconstructed from  identified in the independent analyses of the disc and halo (solid lines), as well as the joint disc-halo decomposition (dashed lines), we find the same features are identified in both the combined and independent analyses: the curves in the upper panel of Figure~\ref{fig:FigF} are unchanged whether the decomposition is performed on a per-component basis, or jointly.
This implies that the same principal component can describe the evolution in both the disc and halo, and that the signal is strong enough in both components to be identified in per-component analyses.
This is a strong indication of a correlated multi-component signal. 
In general the same features will not be recovered from combined analysis of different components because the {\it inter}-component decomposition need not match the {\it intra}-component decomposition. 
In contrast, our joint analysis finds a single PC group may be used to reconstruct the modes in {\it both} the disc and the halo, identifying them as a mutual mode.

For both modes, we can examine and compare timescales and amplitudes to try to understand the driver of the evolution. 
Comparing between components, the feature strength is higher in the halo at earlier times in each mode (of order 1\% density contrast in the halo, but well below that in the disc), implying that the halo is responsible for starting each mode at large radii (compare Figures~\ref{fig:FigC} and \ref{fig:FigE}).
For the growing Mode 1, estimating the growth rate from the modulus of the coefficients at early times also reveals the growth of the halo feature to be twice that of the disc. 
The saturation point of the halo is also measurably earlier than the disc ($T=2.2$ Gyr in the halo vs $T=3.2$ Gyr in the halo). 

The comparison of the disc and halo features in the previous paragraph suggest that the modes may arise from a fundamental dynamical property of the halo component. 
Figure~\ref{fig:FigE} shows snapshots of the halo feature during the simulation at times corresponding to Figure~\ref{fig:FigC}. 
The features are both slow retrograde pattern which build and/or damp over time. 
They bear hallmarks -- a slow dipole pattern at relatively large scales -- of the weakly damped $l=1$ modes in spherical systems that have been studied in using linear perturbation theory. 
These were first identified by \citet{Weinberg...m1...1994}, and later additionally reported by \citet{Heggie...l1...2020}, \citet{Fouvry...l1...2021}, and \citet{Weinberg.dipole.2022}. 

{\it We conclude that BFE+mSSA has allowed us to detect and characterise slow, secular evolution of our isolated simulated galaxy due to the nature of the underlying equilibrium.}  

\subsubsection{Key insights}

The results in this section provide additional illustrations of the ability of BFE+mSSA to separate  evolutionary pathways in a single harmonic and to detect coupling across components.

Most significantly, BFE+mSSA allowed the detection of slow, low-level secular evolution in our simulation that had been predicted in analytic work, \citep{Weinberg...m1...1994, Fouvry...l1...2021} and recently observed in star cluster and dark-matter-halo-only simulations \citep{Heggie...l1...2020, Weinberg.dipole.2022}. The analytic work suggests that spherical systems, such as dark matter halos, generically exhibit dipole point modes. The common existence of these modes has important implications for understanding lopsidedness in galaxies: the halo and disc mutually open dynamical avenues that cannot be taken by either component independently; therefore many dynamical features are simply inexplicable without an understanding of the interplay between components. However, making a clear connection between the theory and observed galaxies has been hampered by the technical challenge of applying analytic work  to multi-component systems. Moreover, while numerical simulations routinely represent multiple component systems, the description  of the results is typically limited to visualisations and statistical analyses that can only qualitatively be connected to dynamical drivers. 

BFE+mSSA has bridged this gap by clearly showing an $l=1$ mode in our simulated halo driving lopsidedness in our simulated disc. These results speak to the promise of BFE+mSSA for forging the missing connection between theory, simulations, and observations needed to interpret galactic properties in terms of our fundamental dynamical understanding secular evolution.

\subsection{Fluctuations and other uninterpretable features} 
\label{subsec:Nullity}

In the two previous sections, we identified interpretable signals in various harmonics of both the disc and halo coefficients in groups of low-order PCs using BFE+mSSA.
However, inspection of the last column of Tables \ref{tab:disequilibrium} and \ref{tab:realdynamics} shows that these PC groups only contain a fraction of the total singular values (which are normalised to total unity): most of the groups represent less than 20 per cent of the variance in the coefficients being analysed\footnote{The exception are some of the PCs associated with the monopole, which encode the equilibrium. These PCs are responsible for upwards of 60 per cent of the singular value signal, cf. Table~\ref{tab:disequilibrium}.}. 
The rest of the signal spread over many (many!) higher-order PCs with lower SVs.  These are PCs with very weak self-gravity.     
We refer to these remaining terms as the {\it nullity}, owing to its uninterpretable nature: it will contain numerical noise, but may also contain signals too weak to be included in our analysis.

\begin{figure*}
  \centering
  \includegraphics[width=1.\linewidth]{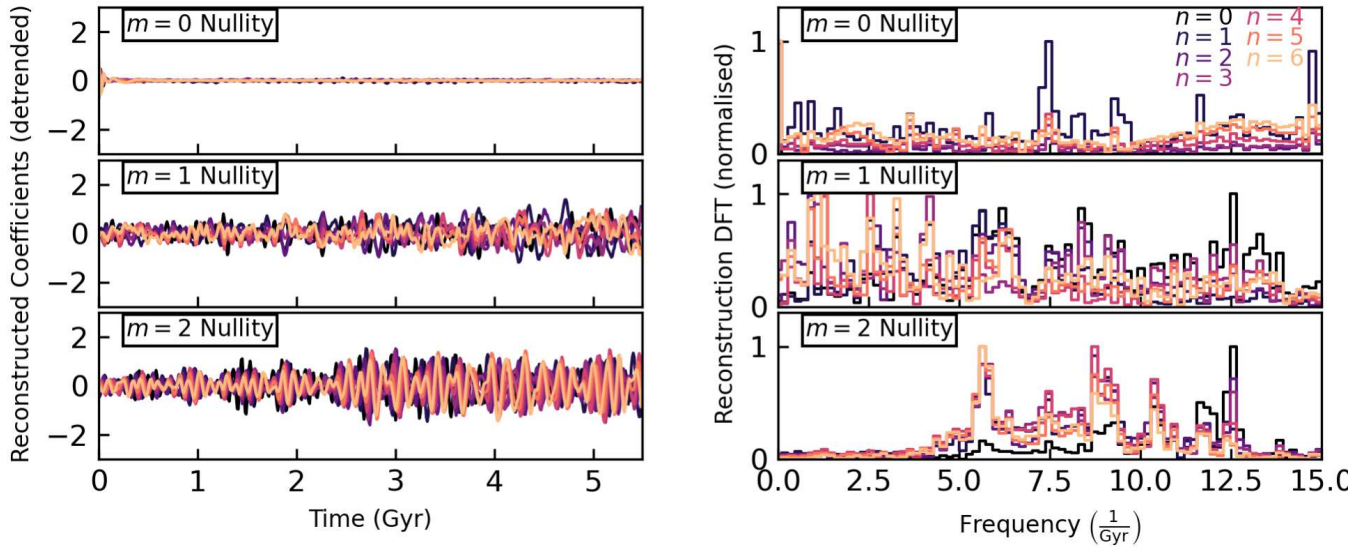}
  \caption{An analysis of the content in the nullity for $m=0$ (upper panels), $m=1$ (middle panels), and $m=2$) lower panels. The left panels show the reconstructed nullity coefficient amplitudes over time for $m=0,1,2$ (top to bottom). The right panels show power spectrum of the reconstructed nullity coefficients for each harmonic. Both $m=0$ and $m=1$ show no discernible signals. The $m=2$ harmonic shows some periodicity, but the power spectrum suggests the frequencies are broad and not strongly coherent. Therefore, we are confident that we are not throwing away interpretable signal in the nullity in any harmonics. These reconstructions may be compare to the unprocessed coefficients, Figure~\ref{fig:FigA}, for a quantitative analysis of what signals are part of coherent signal groups.}
  \label{fig:FigJ}
\end{figure*}

To understand the properties of the nullity, we collect all uninterpretable PCs for a given mSSA decomposition and analyse their reconstructions, summarising the results for low-order disc harmonics in Figure~\ref{fig:FigJ} and for all decompositions in Table~\ref{tab:nullity}. Figure~\ref{fig:FigJ} shows the reconstructed coefficients and corresponding power spectrum for the PCs assigned to the nullity for low-order disc harmonics. 
Comparing this to the corresponding Figures~\ref{fig:FigG} and \ref{fig:FigB} for lower order PCs, the difference is clear. The bottom panels for the $m=2$ nullity do have hints of a signal in the form of low-level systematic evolution in the left hand panel and some clear peaks in the right panel. We discuss future strategies to hunt for weak signals in Section~\ref{significance}.
 However, in general, there is a lack of periodic or systematic evolution in the left hand panels and flat spectra of frequencies in the right hand panel, characteristic of noise. 
A comparison of the contrast columns of Tables~\ref{tab:nullity}, \ref{tab:groups} and \ref{tab:realdynamics} shows that
the fluctuations in the surface density derived from the nullity are mostly stronger than the coherent signals in this particular simulation: our BFE+mSSA analysis has supported insights that would otherwise be inaccessible.

% Summary table of the nullity features, for estimates of missed features
\begin{table*}
\centering
  \begin{tabular}{|c|c|c|c|c|c|c|}
  mSSA                   & PCs     & DFT peak     & contrast  & SV                \\
  decomposition          &         & (Gyr$^{-1}$) & $(R<R_d)$ & fraction          \\
    \hline
    \hline
    disc $m=0$           & 6+      & -            & 0.005   & 0.275               \\
    disc $m=1$           & 6+      & -            & 0.008   & 0.678               \\
    disc $m=2$           & 2+      & -            & 0.017   & 0.799               \\
    disc $m=3$           & 2+      & -            & 0.012   & 0.936               \\
    disc $m=4$           & 2+      & -            & 0.010   & 0.914               \\
    disc $m=5$           & 2+      & -            & 0.006   & 0.954               \\
    disc $m=6$           & 2+      & -            & 0.005   & 0.964               \\
    disc $m=1,3,5$       & 2+      & -            & 0.027   & 0.933               \\
    disc $m=2,4,6$       & 2+      & -            & 0.037   & 0.928               \\
    halo $l=0$           & 6+      & -            & -       & 0.028               \\   
    halo $l=1$           & 6+      & -            & -       & 0.693               \\   
disc $m=0$, halo $l=0$   & 6+      & -            & 0.021   & 0.130               \\     
disc $m=1$, halo $l=1$   & 8+      & -            & 0.010   & 0.667               \\       
disc $m=1,2$, halo $l=1$ & 4+      & -            & 0.012   & 0.801               \\  
\end{tabular}
  \caption{Summary of principal components assigned the nullity in our decompositions. We refer to each collection of PCs here as the `Nullity', rather than a PC group. Disc harmonics are denoted with $m$, halo harmonics are denoted with $l$. Columns are the same as in Table~\ref{tab:disequilibrium}.}
  \label{tab:nullity}
\end{table*}

\section{Looking Ahead}
\label{sec:LookingAhead}

\subsection{Essential Future Work - assessment of weak feature significance} 
\label{significance}

Our analyses of simulations of bar formation \citep{Weinberg...mSSA...2021,Petersen...exp...2022} and an isolated disc galaxy (this paper) amply illustrate the facility of BFE+mSSA to learn about both significant and expected as well as subtle and unanticipated dynamical evolution. 
The results are very promising for general applications to a wide variety of dynamical systems.
However, our work so far has been involved close supervision of BFE+mSSA to both interpret and understand the significance of what features it has identified. 

In particular, the interpretative ambiguity we encountered in the higher order terms in this paper outlines the current limit of BFE+mSSA.
This limit motivates the need for a rigorous statistical analysis of significance for mSSA-identified signals.
Many of the well-known approaches from statistical analysis would be suitable for this purpose.
For example, let us take the hypothesis that the signal observed at \(m=2,3,4\) is consistent with background noise as a test case. 
That is, our null hypothesis is that our simulation can generate with the same properties of the signal in question without inherent self gravity.
To do this, we need to generate a simulation with the same noise spectrum as the full simulation but without any self-gravitating features on the spatial and temporal scales of our putative signal.
Let us assume that we know how to perform such simulations (we propose an \EXP{}-enabled approach below). 
An ensemble of these null-hypothesis simulations can be run and analysed using mSSA.
From the ensemble of simulations, one may construct prediction intervals for singular values under the null hypothesis.
Then, if the singular value corresponding to the signal in question is beyond the prediction intervals, the corresponding principal component is considered significant. 
In such a case, the signal can be reliably reconstructed. 
This approach is often called Markov Chain SSA (MC-SSA, see \citealt{Allen.Smith:1996}).

Analyses of this sort are particularly well-suited to the \EXP{} framework described in \cite{Petersen...exp...2022}.  
We can use the mSSA analysis to construct a realistic reconstruction of the coefficients series from the self-gravitating simulation \emph{without} the self-gravitating features of interest by removing the groups corresponding to the signal in question.
In the study presented here, this would be akin to retaining only the nullity reconstructions of the coefficients.
We can generate new coefficient series from an autoregressive model\footnote{Autoregressive noise models are typically used for null hypotheses in MC-SSA because SSA provides good estimates for frequencies and exponential factors processes generated by the related linear recurrence relations \citep[Section 3]{Golyandina.Zhigljavsky:2013}.} consistent with the coefficient covariance from the mSSA reconstruction. 
Then, \EXP{} allows initial potential fields from the reconstructed coefficients to be replayed for a new ensemble of particles with very little computational effort.
The resulting expansion coefficient series are gathered automatically for analysis by mSSA, and can be analysed for significance of detected features.
A detailed description of the MC-SSA approach in the \EXP{} context will be described in a later contribution.

\subsection{Prospects for applications to simulations} 
\label{sec:prospects}

Despite the limitations, there are multitude of prospects for immediate, supervised applications of BFE+mSSA to simulations of galaxies, whether isolated, interacting or evolving in the full cosmological context.
\begin{description}
\item[\underline{Dynamical analyses of simulations of galactic evolution.}]  Recent surveys \citep{Majewski...APOGEE...2017,Steinmetz...RAVEdr6...2020,Gaia...DR3...2022} demonstrate that the Milky Way continues to evolve through satellite interaction. 
$N$-body simulations have explained some key observational signatures \citep{Laporte...SgrGaia...2019,Petersen...reflexprediction...2020,Garavito-Camargo...wakes...2021,Vasiliev...Sgr...2021,Hunt...snails...2022}.
However, interpretation of these simulations is challenging since many actors contribute simultaneously. 
The BFE+mSSA knowledge discovery approach is capable of separating, characterising and dissecting the signatures of the mutual interactions of each component in simulations by separating features by correlating temporal and spatial scales non-parametrically.
BFE+mSSA promise detailed predictions and identification of features in current stellar data sets \citep[see][for some recent results]{Petersen...reflexmeasurement...2021,Garavito-Camargo...poles...2021,Lilleengen...streams...2022} and confident mapping the the dark matter halo's global structure and {\it distortions} to that structure.  This goal was unimaginable even 5 years ago.
\item[\underline{Structural characterisation and correlation of fields.}]  This paper demonstrated the discovery of two-dimensional features in disc density resulting from internal (disequilibrium-related) dynamics and halo interactions.  
However, BFE+mSSA can be applied to any field in any number of dimensions.  
For example, \citet{Weinberg...mSSA...2021} illustrated a three-dimensional disc BFE.  
The \EXP\ library already enables joint BFE+mSSA investigations of any number of three-dimensional density and potential fields.  These may be augmented by kinematic fields as in \citet{Weinberg...mSSA...2021} or some other field such as star formation rates and implied local metallicity. 
If the additional fields encode spatial information (e.g. they are BFE coefficients or even radial and azimuthal bins), their temporal and spatial scales will be correlated with the density and potential fields.  
The BFE+mSSA can adapt to new observational tools and \emph{windows} as new surveys become available.
\item[\underline{Understanding of noise.}]  There have been many years of debate on the effect of noise in conclusions drawn from dynamical simulations, from bar-halo interactions \citep{Weinberg.Katz..nbody..2007}, through dynamical friction \citep{Weinberg..dynamicalfriction..2001}, to satellite disruption \citep{Errani...resolution...2020}. 
BFE+mSSA clearly separates the correlated, quasi-periodic signals resulting from dynamical interaction and coupling from the fluctuating forces resulting from finite particle number stochastic effects.  
We expect that couplings in orbital dynamics have frequencies near or smaller than the characteristic orbital frequencies.  
Since the individual PCs describe the temporal behaviour of components assigned to the noise field and the power spectrum describes their characteristic frequencies, mSSA provides a natural classification of signal and noise. 
Investigations of test-particle orbits with and without the noise component provide a diagnostic tool for the reliability of features in simulations and the role of fluctuations more generally. 
\end{description}

\section{Conclusions}
\label{sec:Conclusions}

\subsection{Near-equilibrium evolution: the importance of multi-component modes} 
\label{subsec:NearEquilibriumEvolution}

We applied BFE+mSSA to a simulation of an isolated Milky Way like galaxy. 
The BFE+mSSA combination allows us to \emph{automatically} identify the main features in the model galaxy {\it and their origins}.
Most remarkably, BFE+mSSA achieved this in the challenging case of an isolated, multi-component galaxy that had specifically been constructed to {\it not} evolve and where the dynamical signatures were below the level of the noise. Our work complements a prior investigation \citep{Weinberg...mSSA...2021} which used BFE+mSSA to characterise significant evolution of known nature in a simulation which formed a galactic bar.

In our near-equilibrium model, we identified -- for the first time -- two multi-component (disc-halo) dipole point modes (Figures~\ref{fig:FigB}-\ref{fig:FigF}) which evolve over time (one growing, one damped; Table~\ref{tab:realdynamics}). This discovery is enabled by the BFE+mSSA methodology; such dynamical effects are at a level such that other methods, such as Fourier analyses, will not be able to recover the signals.
Halo modes are expected from linear perturbation theory \citep{Weinberg...m1...1994,Fouvry...l1...2021}, and are observed in simulations of star clusters \citep{Heggie...l1...2020} and dark matter halos \citep{Weinberg.dipole.2022}, but the coupling of a spheroid to a disc has not been discovered to date.
The BFE+mSSA methodology makes the identification of point modes straightforward, and provides several avenues for corroboration. 
We employed several different mSSA decompositions to validate our findings. 
The existence of point modes in these isolated simulations demonstrates the fundamental contribution of component interactions to the dynamical evolution of galaxies.
We expect that the existence of such multi-component disc-halo modes is a generic feature of such systems, possibly including the Milky Way.
These modes likely have influence over the structural evolution of disc galaxies.
For instance, our results immediately suggest that low dipole modes will be most detectable at large radii (e.g. $R>20$ kpc in the Milky Way), where density contrasts can exceed 10 per cent relative to a smooth disc.

In addition to point modes, we identified the long-lived results of initial conditions disequilibrium, resulting from individual halo and disc disequilibrium features.
Starting with rings encoded in the monopole $m=0$ coefficients, we found correlations with many other harmonics, including a persistent $m=2$ signal.
We uncovered an aphysical `settling' of the halo in response to the presence of the disc at the outset of the simulation.
Future work modelling idealised galaxies must take care to ensure that disequilibria in the initial conditions, and the resulting persistent features, are not treated as real dynamics.

Finally, we quantified the remaining signal that we classified as the nullity, and put limits on the magnitude of unexplained surface density fluctuations in the disc. The desire to push even deeper in the decomposition of simulations motivates the essential future work, but also inspires prospects for future applications.

\subsection{Dynamical Data Mining as the future of Galactic Dynamics}
\label{subsec:DynamicalDataMining}

Galactic Dynamics is a mature field with elegant descriptions of equilibrium systems, estimates for scales of the processes involved in the interactions that are known affect them and sophisticated analytic methods that describe evolution in the linear regime. 
While we can understand the basic governing principles with detailed mathematical models from Hamiltonian perturbation theory, the inter-component and environmental interactions that produce this morphology are hard if not impossible to study from modal analysis alone.
BFE methods both underpin our dynamical data mining technique and are often used in analytic perturbation work. 
Thus they provide a natural bridge between theoretical work and numerical simulation.  
The combination of BFE representation of the possibly unknown dynamics in simulations with a machine-based knowledge acquisition tool such as mSSA allows for identification of couplings that may be too hard to predict otherwise.  
This natural synergy between mathematical theory and simulation is the main motivation for our approach.
Series of BFE can also be used to characterise {\it observed} fields in galaxies. 

A galaxy's picturesque morphological structure is a historical summary of its evolution.  
Cosmological predictions for the frequency of galactic interactions explain the abundant signatures of disequilibrium observed.  
The detail of our picture of disequilibrium is rapidly advancing, in terms of resolution in the Galaxy \citep[e.g.][]{Hunt..gaiasnails..2022} in galaxies and occurrence rate in others \citep[e.g.][]{Pearson..externalstreams..2022}.
Simulated realisations of galaxies in disequilibrium are similarly advancing in resolution and scale.
However, the tools to take full advantage of this twin onslaught of data, simulated and real, are currently lacking. 
Such tools must be capable of modelling galaxies in disequilibrium, make quantitative and dynamically meaningful connections between simulated and observed galaxies and connect with analytic work in the linear regime. 

As outlined in Section \ref{sec:prospects}, our results suggest the tremendous promise of BFE+mSSA for the field of Galactic Dynamics, with a myriad of envisioned applications. 
Many of these applications can be undertaken now by adopting the supervised learning approach to using BFE+mSSA. These include detailed analyses of galactic components for galaxies in both isolated and cosmological settings. 

Nor is there any reason to limit the BFE+mSSA analysis to galaxies. BFE+mSSA can be used for the characterisation and dynamical evolution of self gravitating, interacting systems more generally and in any context, from binary asteroids in the solar system \citep{Quillen..asteroids..2022}, through proto-planetary discs \citep{Cadman..protoplanetary..2021}, to nuclear star clusters in the centres of galaxies \citep{Fouvry..starclusters..2022} .

The remaining and key challenge to be solved is to understand how to confidently assess the significance of all the features that BFE+mSSA recovers in an {\it unsupervised} way. 
Once this is developed it will be possible to broadly apply BFE+mSSA to large samples of systems: both simulated and real.

\section*{Acknowledgements}
We warmly thank Chervin Laporte for sharing the initial conditions from his simulation with us.
We acknowledge support from the Center for Computational Astrophysics (CCA) at the Flatiron Institute in the form of access to their computational resources which allowed us to create our simulation and generate the associated data.
In addition, we thank CCA leadership and staff for hosting the Beyond-BFE collaboration meetings.
We thank members of the B-BFE collaboration and the Dynamics Group at CCA for numerous conversations during development of this paper.
MSP's contributions were partially supported by grant Segal  ANR-19-CE31-0017 of the French Agence Nationale de la Recherche as well as a UKRI Stephen Hawking Fellowship.
KVJ and AJ's contributions were supported by NSF grant AST-1715582.

\section*{Data Availability}

The code, data, and simulation used to generate the results in this article will be made available upon reasonable request to the appropriate author.

\bibliographystyle{mnras}
\bibliography{mssa}

\appendix

\section{All mSSA combinations tested}
\label{appendix:allfeatures}

% definition of evolution options:
% `none` if early and late times are the same, or undetermined
% `decaying` if early time is obviously stronger than late
% `growing` if late time is stronger than early

% If we have a pure m=1 component and we slightly shift to a new  origin, the mapping required to translate the to the new center requires pieces from all the harmonics.   This is the shift operator for polar harmonics.  It's most often used for spherical harmonics.   I don't know if I've seen the exact formula for polar, but if it's anything like spherical, odds tend to involve other odd terms preferential.
% So I would say that this is a 'projection' , sure.  But  'alias' would be a misnomer in a paper that talks about  time series.   I wonder if 'projection' needs a quick description in Appendix A?

% Summary table of the features we will discuss.
\begin{table*}
\centering
  \begin{tabular}{|c|c|c|c|c|c|c|c}
                       &         &         & Fourier      & visual    &             & singular            &                  \\
    mSSA               &         &         & peak         & contrast  & evolution   & value               &                  \\
decomposition          & Group   & PCs     & (Gyr$^{-1}$) & $(R<R_d)$ & type        & fraction            & interpretation   \\
    \hline
    \hline
    \multicolumn{8}{c}{disc-only decompositions}\\
    \hline
\multirow{3}{*}{$m=0$} & 1       & 0,1     & 0.2          & 0.031   & slow decay    & 0.641               & phase mixing of halo initial conditions\\
                       & 2       & 2,3,4,5 & 6.4          & 0.006   & fast decay    & 0.084               & phase mixing of disc initial conditions \\
                       & nullity & 6+      & -            & 0.005   & no evolution  & 0.275               & - \\
    \hline
\multirow{4}{*}{$m=1$} & 1       & 0,1     & 0.6          & 0.007   & slow growth   & 0.201               & coupling with halo \\
                       & 2       & 2,3     & 1.7          & 0.003   & slow decay    & 0.064               & coupling with halo \\
                       & 3       & 4,5     & 6.9          & 0.002   & fast decay    & 0.057               & phase mixing of disc initial conditions \\
                       & nullity & 6+      & -            & 0.008   & no evolution  & 0.678               & - \\
    \hline
\multirow{2}{*}{$m=2$} & 1       & 0,1     & 6.6          & 0.006   & slow growth   & 0.201               & phase mixing of disc initial conditions \\
                       & nullity & 2+      & -            & 0.017   & no evolution  & 0.799               & - \\
    \hline
\multirow{2}{*}{$m=3$} & 1       & 0,1     & 13.8         & 0.005   & slow growth   & 0.064               & projection of $m=1$ Group 1 \\
                       & nullity & 2+      & -            & 0.012   & no evolution  & 0.936               & - \\
    \hline
\multirow{2}{*}{$m=4$} & 1       & 0,1     & 14.2         & 0.004   & slow growth   & 0.086               & projection of $m=2$ Group 1 \\
                       & nullity & 2+      & -            & 0.010   & no evolution  & 0.914               & - \\
    \hline
\multirow{2}{*}{$m=5$} & 1       & 0,1     & 23.1         & 0.001   & slow growth   & 0.046               & projection of $m=1$ Group 1 \\
                       & nullity & 2+      & -            & 0.006   & no evolution  & 0.954               & - \\
    \hline
\multirow{2}{*}{$m=6$} & 1       & 0,1     & 20.2         & 0.001   & slow growth   & 0.036               & projection of $m=2$ Group 1 \\
                       & nullity & 2+      & -            & 0.005   & no evolution  & 0.964               & - \\
    \hline
\multirow{2}{*}{$m=1,3,5$} & 1   & 0,1     & 0.6          & 0.008   & slow growth   & 0.067               & projection of $m=1$ Group 1 \\
                       & nullity & 2+      & -            & 0.027   & no evolution  & 0.933               & - \\
    \hline
\multirow{2}{*}{$m=2,4,6$} & 1   & 0,1     & 6.6          & 0.010   & slow growth   & 0.072               & projection of $m=2$ Group 1 \\
                       & nullity & 2+      & -            & 0.037   & no evolution  & 0.928               & - \\
    \hline
    \hline
    \multicolumn{8}{c}{halo-only decompositions}\\
    \hline
\multirow{3}{*}{$l=0$} & 1       & 0,1,2,3 & 0.4          & -       & slow decay    & 0.944               & phase mixing of halo initial conditions \\
                       & 2       & 4,5     & 6.6          & -       & fast decay    & 0.028               & phase mixing of disc initial conditions \\
                       & nullity & 6+      & -            & -       & no evolution  & 0.028               & - \\   
    \hline
\multirow{3}{*}{$l=1$} & 1       & 0,1,2,3 & 0.4          & -       & slow growth   & 0.272               & weakly self-gravitating mode \\
                       & 2       & 4,5     & 1.5          & -       & slow decay    & 0.035               & weakly self-gravitating mode \\   
                       & nullity & 6+      & -            & -       & no evolution  & 0.693               & - \\   
    \hline
    \hline
    \multicolumn{8}{c}{disc-halo decompositions}\\
    \hline
\multirow{3}{*}{$m=0$,$l=0$} & 1 & 0,1,2,3 & 0.2          & 0.054   & slow decay    & 0.832                & phase mixing of halo initial conditions \\
                       & 2       & 4,5     & 6.6          & 0.007   & fast decay    & 0.037                & phase mixing of disc initial conditions \\
                       & nullity & 6+      & -            & 0.021   & no evolution  & 0.130                & - \\     
  \hline
\multirow{4}{*}{$m=1$,$l=1$} & 1 & 0,1,2,3 & 0.6          & 0.008   & slow growth   & 0.244                & weakly self-gravitating mode and coupling \\
                       & 2       & 4,5     & 1.5          & 0.003   & slow decay    & 0.048                & weakly self-gravitating mode and coupling \\
                       & 3       & 6,7     & 6.9          & 0.002   & fast decay    & 0.040                & phase mixing of initial conditions and coupling \\
                       & nullity & 8+      & -            & 0.010   & no evolution  & 0.667                & - \\       
    \hline
\multirow{3}{*}{$m=1,2$,$l=1$} & 1 & 0,1   & 0.6          & 0.010   & slow growth   & 0.117                & weakly self-gravitating mode and coupling \\
                       & 2       & 2,3     & 6.6          & 0.003   & fast decay    & 0.082                & phase mixing of disc initial conditions and coupling \\
                       & nullity & 4+      & -            & 0.012   & no evolution  & 0.801                & - \\  
\end{tabular}
  \caption{Summary of different modes identified in our MSSA decompositions. Disc harmonics are denoted with $m$, halo harmonics are denoted with $l$. Disc feature strengths are reported in surface density to give a measure of `visual contrast', defined as $\max\left(|\Delta_\Sigma|\right)$ within a disc scale length (see equation~\ref{eq:contrastequation}). Contrasts have an approximate error of 0.001, estimated from grid size adjustments. Owing to simulation sampling rates (0.01 Gyr), the DFT peak is only accurate to 0.1. The group name can be derived for each row by combining the decomposition harmonic(s) and concatenating with the group number. For example, the name of the first row would be Group $m$0-1.}
  \label{tab:groups}
\end{table*}

Table~\ref{tab:groups} summarises all tested combinations in mSSA. All tests include all radial orders, i.e. $n\in[0,6]$. All mSSA decompositions use a window length $L=250$, which is approximately half of the input time series ($N=549$). This creates sets of 250 PCs. The PCs are sorted by singular value magnitude, which does not guarantee that PCs with physical similarity are consecutive. Therefore, we determined the grouping of the PCs through direct examination, which in practice was straightforward. The information in this table is repeated from Tables~\ref{tab:disequilibrium}-\ref{tab:nullity} in text, consolidated here and reorganized by mSSA decomposition for ease of comparison.

\section{Basis function expansion details}
\label{sec:bfedetails}

In this Appendix, we briefly describe the basis function expansion technical implementation for completeness and clarity. Full details may be found in \citet{Petersen...exp...2022} for the spherical expansions, and in \citet{Weinberg...mSSA...2021} for the Fourier-Laguerre disc expansion.

\subsection{Spherical expansions: Empirical Orthogonal Functions}

For an initially spherically-symmetric model (such as a dark matter halo or stellar bulge) we use spherically symmetric basis functions derived using the machinery in \EXP{}. The potential and density take the form
\begin{equation}
\phi_{nlm} = \phi_0(r)u_{nl}(r) Y_{lm}(\theta,\phi)
\end{equation}
\begin{equation}
\rho_{nlm} = \rho_0(r) u_{nl}(r) Y_{lm}(\theta, \phi)
\end{equation}
where $u_{nl}$ are eigenfunctions determined by \EXP{}, $Y_{lm}(\theta,\phi)$ are the usual spherical harmonics\footnote{For ease of numerical implementation, \EXP{} uses the real spherical harmonics: $Y_{lm,\cos}=\frac{1}{2}\left(Y_{lm}+Y_{l-m}\right)$ and $Y_{lm,\sin}=\frac{1}{2i}\left(Y_{lm}-Y_{l-m}\right)$.}, and $\rho_0(r)$ and $\phi_0(r)$ are input unperturbed model density and potential. The model density and potential are typically chosen to be the initial conditions. The functions $u_{nl}(r)$ are eigenfunctions of a the Sturm-Liouville equation. Each function is a solution to the Poisson equation and has $n$ nodes with increasingly tighter spacing (as $n$ increases). The $u_{00}(r)$ function is a constant, which makes the $\ell=0,~n=0$ potential and density terms exactly proportional to the input model. Other terms are then perturbations on top of the model potential and density. The functions are biorthogonal, satisfying two conditions:
\begin{equation}
\int\text{d}^3\mathbf{v} \;\; \phi_{nlm} \; \rho_{n'l'm'} \propto \delta_{nn'}\delta_{ll'}\delta_{mm'}
\end{equation}
\begin{equation}
\nabla^2 \phi_{nlm} = 4\pi G\rho_{nlm}.
\end{equation}
The derivation of these functions is described in \citet{Petersen...exp...2022}.
For the purposes of this work, the BFE representation of density is given by the projection of the coefficients onto the BFE:
\begin{equation}
\hat{\rho}_{\rm halo}(r,\theta,\phi;t) = \sum_{l}\sum_{m}\sum_{n}A_{lmn}(t)\rho_0(r) u_{nl}(r) Y_{lm}(\theta, \phi)
\label{eq:sphericalcoefficientprojection}
\end{equation}
where $A_{lmn}(t)$ is the coefficient amplitude for a given function indexed by $(l,m,n)$, and possibly is a function of time. An analogous expression may also be written for the potential. 
The $\hat{\cdot}$ notation indicates that the quantity is reconstructed from the coefficients.

\subsection{Disc surface density expansion: Fourier-Laguerre} \label{discbfe}

A spherical expansion is not appropriate for a strongly flattened stellar disc. While \EXP{} supports a three-dimensional empirical orthogonal function basis that may be used to represent the potential and density, we choose in this work to project the disc to the two-dimensional plane. We work in polar coordinates $R$ and $\phi$ for a two-dimensional expansion. For the radial coordinate we use the Laguerre basis functions described in \citet{Weinberg...mSSA...2021}, which were created to match the exponential profile of a typical stellar disc. This choice minimises the number of required functions. The Laguerre polynomials are defined as:
\begin{equation}
G_n(R)=\frac{1}{a\sqrt{n+1}}~\exp\left(-\frac{R}{a}\right)~L_n^1\left(\frac{2R}{a}\right),
\end{equation}
where $L_n^1$ is the associated Laguerre polynomial of order 1 and degree $n$, and $a$ is the scale length of the disc. Then $G_0(R)=\frac{2}{a}e^{-R/a}$ closely approximates the disc density and the majority of the reconstruction power resides in a single term. Moreover, the polynomials satisfy the the orthogonality condition
\begin{equation}
\int G_n(R)G_{n'}(R) R dR = \delta_{nn'},
\end{equation}
and thus $G_n(R)$ can be used to reconstruct the radial structure of the disc with a small number of expansion terms.

The azimuthal dependence is described with a Fourier series. Combining this with the radial Laguerre basis functions we obtain a set of two-dimensional basis functions $G_n(R)\cos{m\phi}$ and $G_n(R)\sin{m\phi}$. We call these Fourier-Laguerre functions. The coefficients are naturally determined by the projection of $f(R,\phi)$ into the basis functions. In our case we are accumulating discrete particles so rather than a continuous integral we obtain the discrete summations:
\begin{equation}
C_{mn} = \frac{1}{2\pi}\sum_i f(R_i, \phi_i)e^{im\phi_i} G_n(R_i)
\end{equation}
where $(R_i,\phi_i)$ is the position of particle $i$. The reconstruction of the surface density field is then
\begin{equation}
\hat{\Sigma}_{\rm disc}(R, \phi;t) = \sum_m\sum_n C_{mn}(t)e^{im\phi}G_n(R),
\label{eq:fourierlaguerrereconstruction}
\end{equation}
where the coefficients $C_{mn}(t)$ may have some time dependence.

\section{Multichannel Singular Spectral Analysis}
\label{sec:mSSA}

Singular Spectral Analysis (SSA) is a method of non-parametrically decomposing a time series into a sum of components that ideally capture different aspects of the series. Multichannel Singular Spectral Analysis (mSSA) extends SSA to include multiple series, such that one may identify coherent signals between series. A full description and pedagogical examples may be found in \citet{Weinberg...mSSA...2021}. In this appendix, we briefly describe the important elements of SSA and describe the implementation used in this paper. 

\subsection{A conceptual introduction to mSSA}
\label{sec:mssabrief}

In this section, we present a brief and intuitive introduction to mSSA through the conceptual relationship with principal component analysis (PCA).
PCA characterises correlations within a data set by transforming to coordinates  where most of the correlation is represented in a small number of dimensions in \emph{space}. 
mSSA itself is a generalisation of PCA: mSSA performs computations analogous to PCA on a {\it grand trajectory matrix} which is constructed to represent different variations in {\it time} intervals. 
The extension of PCA to multiple times allows for the decomposition of different series of samples of variables over time. 
In our case, the variables are the individual BFE coefficients for each snapshot in the simulation. 
We can then simultaneously decompose structure in space and time.  
We refer the interested reader to \citet{Golyadina...ssabook...2001} for a more thorough description of SSA.

\subsubsection{The goals of Principal Component Analysis}

The goal of PCA is the reduction of the data set's dimensionality
while retaining as much as possible of the variation present in the
data set.  This is achieved by a linear transformation or
\emph{rotation} to a new set of axes, the principal components
(PCs), which are uncorrelated and ordered by their contribution to the
total variance. For a brief introduction, consider a data sample of \(N\)
random variables with \(M\) channels (or dimensions). PCA is mathematically equivalent
to the eigen-analysis of the covariance matrix:
\begin{equation}
  \boldmatrix{C} \propto \boldmatrix{X}^\intercal\cdot\boldmatrix{X}
  \label{eq:covar0}
\end{equation} 
where \(\boldmatrix{X}\) is a \(M\times N\) matrix whose \((i, j)\)th
element is the data with zero-mean and unit variance for each of the
\(M\) channels: \((x_{ij} - \mu_i)/\sigma_i\) where \(\mu_i,
\sigma_i^2\) is the mean and variance for channel \(i\), respectively\footnote{The covariance matrix is often normalised in PCA as \(\boldmatrix{C} = \frac{1}{d}\boldmatrix{X}^\intercal\cdot\boldmatrix{X}\), where if the mean is determined from the data, i.e., the sample variance, we have
\(d=N-1\).}.
The sums over covariance between channels is effected by the matrix
multiplication in equation (\ref{eq:covar0}). Then, the leading eigenvector or PC is the direction that maximises the variance.  
The next PC is the direction that maximises the variance, uncorrelated to the first, and so on.  

\subsubsection{Relationship of mSSA and Principal Component Analysis}\label{subsec:mssa}

For our analysis, we have time series of \(N\) samples for
each of \(M\) BFE coefficients that have been reduced to zero mean and
unit variance: \(\{a_{i,j} : i=1,\ldots,M; j=1,....,N\}\).  
Each of the \(N\) samples corresponds to a phase-space snapshot in our simulation, providing a two-dimensional grid in coefficient channel (the \(i\) index) and time (the \(j\) index). 
A PCA analysis of this matrix, as described above, identifies the sets of weighted coefficients (the principal components) which contain the highest proportion of the variance in each of the \(N\) temporal views. 
Each of these sets represents  a coherent {\it spatial} pattern that is present in the data.

The mSSA algorithm adds the ability to simultaneously find dominant shapes in space that also evolve similarly in time\footnote{SSA is a unique case where \(M=1\). That is, we only have one input channel. In this work, we advocate for use of mSSA specifically, but one could also apply the discussion here to SSA.}. 
For our case, this is done by constructing a new matrix where each row contains information from a sequence of $L < N$ snapshots rather than a single one, effectively creating a time {\it window}. 
Each of $K=N-L+1$ rows is of  length $M\times L$. 
The $i$th row is a concatenation of the coefficients derived from snapshots $i$ - $i+L-1$, so each contains the same coefficients, but lagged by systematic amounts in time.  
The resulting $(M\times L) \times K$ matrix is known as the {\it grand trajectory matrix} \citep{Ghil...review...2002}. 
Constructing {\it grand covariance matrix} and performing a PCA would now compare not one, but $L$ snapshots at once to find repeating patterns at different lag times, effectively sliding a window of length $L$ over the simulation.
The resulting PCs maximise the variance in the \(K = N-L+1\) overlapping views of the \(M\) time series simultaneously.

For intuition, consider a set of coefficients that all have sinusoidal variation with a single period.  At time lags that are multiples of the period, the variance will be large as the signals reinforce each other coherently.  At incommensurate lags, the covariance will tend to zero.  This allows mSSA to naturally find the coherent temporal signals in the data.   For each oscillatory signal in our series, we will find a pair of eigenvectors that represent the same frequency, just as in Fourier analysis.  A sinusoid was simply an example; the method will work any temporally coherent signal including exponential growth or decay.  mSSA is purely non-parametric in this sense.

\subsection{Details of mSSA implementation}

SSA is principal component analysis (PCA) of sequentially lagged $L$-length windows of a time series (where $L$ is a user-specified {\it window length}). 
The key is that distinct features have different projections into $L$-lagged space and so PCA separates the different features. 
When working optimally on a dynamical system input, the different features will correspond to different dynamical phenomena, such as a galactic bar \citep[as in][]{Weinberg...mSSA...2021}. 
The SSA procedure involves three main steps: embedding, singular value decomposition, and grouping/reconstruction.

\subsubsection{Embedding}

In this step of SSA one forms a matrix which represents the sequence
of $L$-lagged windows of the time series. This matrix is called the
trajectory matrix. Consider an input time series $\vec{s}=\{s_1 \dots s_N\}$ and window
length $L$. The trajectory matrix is formed as
\begin{equation}
  \boldmatrix{T} =
  \begin{bmatrix}
    s_1 & s_2 & \dots & s_{N-L+1} \\
    s_2 & s_3 & \dots & s_{N-L+2} \\
    \vdots&     & \ddots& \vdots \\
    s_L & s_{L+1} & \dots & s_N
  \end{bmatrix}.
\end{equation}
It is common to denote $N-L+1$ as $K$.  The anti-diagonals
of $\boldmatrix{T}$ are equal by construction. Matrices with this property are called Hankel matrices.

\subsubsection{Singular Value Decomposition: Principal Components and Singular Values}
\label{sec:SVD}

After forming the trajectory matrix, the next step is to perform a singular value decomposition (SVD). We begin by inspecting the dimensions of the trajectory matrix, $\boldmatrix{T}\in \mathbb{R}^{L\times K}$. To maximise computational efficiency, we construct the covariance matrix such that it is $\min(K,L)\times\min(K,L)$.

Consider the SVD of the trajectory matrix
\begin{equation}
\boldmatrix{T}=\boldmatrix{U}\boldmatrix{\Lambda}^{1/2}\boldmatrix{V}^\intercal.
\end{equation} The covariance matrix for \(L<K\), \(\boldmatrix{C} = \boldmatrix{T}\boldmatrix{T}^\intercal = \boldmatrix{U}\boldmatrix{\Lambda}\boldmatrix{U}^\intercal\).   When \(K<L\), we can write the covariance as
\(\boldmatrix{C} = \boldmatrix{T}^\intercal\boldmatrix{T} = \boldmatrix{V}\boldmatrix{\Lambda}\boldmatrix{V}^\intercal\). The choice in constructing the covariance matrix is to decrease the computational complexity by choosing the representation with the smallest rank nullity\footnote{The information content is the same in both, since the ``wrong'' choice simply increases the rank of the nullity which has no useful information.}. We may then obtain the SSA {\it principal components} either directly from $\boldmatrix{U}\in \mathbb{R}^{L\times K}$, or as $\boldmatrix{U} = \boldmatrix{T}\boldmatrix{V}$. The SSA empirical orthogonal functions either come directly as $\boldmatrix{V}\in \mathbb{R}^{L\times K}$, or may be computed as $\boldmatrix{V} = (\boldmatrix{U}^\intercal\boldmatrix{T})^\intercal$. The \textit{singular values} are given along the diagonal of $\boldmatrix{\Lambda}$: that is, the $i^{\rm th}$ singular value is $\sigma_i=\sqrt{\lambda_i}$. We sort the singular values from highest to lowest and sort the columns of $\boldmatrix{U}$ and $\boldmatrix{V}$ accordingly.

Given the  $i^{\rm th}$ empirical orthogonal function $V_i$ and a principal component $U_i$, we can form the matrix $\boldmatrix{A}_i=\sigma_iU_iV_i^\intercal$. Summed over all $i$, these matrices are the set of matrices which best approximate $\boldmatrix{T}$ {\it and} have orthogonal column/row spaces. In this sense the SVD of $\boldmatrix{C}$ gives us a set of matrices which each carry independent information. 
We refer to the tuples $(\sigma_i,U_i, V_i)$  as {\it eigentriples}.

\subsubsection{Grouping}
\label{subsubsec:grouping}
 
The various input coefficient series are often not independent.  
This, combined with sampling noise, may lead to information smearing between PCs.
Of practical importance, there are often multiple eigenvectors and PCs that correspond to a particular dynamical signal.  
The mSSA practitioner will need to \textit{group} eigenvectors that describe the same or related signals together.  
For example, even for a pure sinusoidal signal, the basic group will contain two eigenvectors, with identical eigenvalues. The pair together describe the amplitude and phase.

For typical features found in disc galaxies, such as arms and bars, the relevant groups often contain multiple eigenvectors. 
There are a variety of ways to motivate groupings. 
As a starting point, a group of PCs describing a single dynamical features will often have similar eigenvalues.
Next,  plotting
the PCs over time by eye 
often reveals whether their evolution is similar. This can be quantified by (for example) taking a Fourier transform.  
In particular, power spectra can clearly indicate which PCs are describing noise as they will have a broad frequency spectrum. 
The understanding of the noise can also be used to modify the window length to get better noise separation properties.

$\mathbf{w}$-correlation matrices are a good grouping diagnostic \citep{Weinberg...mSSA...2021}. 
These matrices quantify the correlation between PCs and can be used to guide grouping by using the correlation as an indication of which PCs belong in groups. 
Though we have listed two approaches to grouping, it is important to note that these alone do not establish the groups. 
The grouping is  determined by the content of the PCs. We find grouping to be important primarily because misgrouping can spread out the correlated signal and lead to an underestimate of a particular correlated degree of freedom.

As an example, examination of PCs may indicate that two long duration trends are present in the same PC. In this case, the window length could be increased to separate those trends. However, it is also common that PCs may demonstrate the existence of multiple behaviours present in single PCs leading to a modification of the window length in the way described in the previous paragraph to achieve better separation. The converse situation can also occur, where single processes are being split into many PCs, in which case the window length can be shortened to simplify grouping.

The goal of grouping is a partition of the indices $i$ into $m$ distinct subsets such that each set of $m$ subsets corresponds to a distinct feature. 
The trajectory matrix reconstructed from select $i$ indices is
\begin{equation}
    \boldmatrix{\tilde{T}}^k = \sum_{i \in I_m} \sigma_iU_i V_i^\intercal 
\end{equation}
where $I_m$ is a list of $i$ indices in the group. After segmenting the eigen-triples into groups, each $\boldmatrix{\tilde{T}}^k$ ideally corresponds to a distinct feature.

\subsubsection{Coefficient Reconstruction}\label{subsubsec:reconstruction}

Given that the matrices $\boldmatrix{\tilde{T}}^k$ need not be (and almost certainly will not be) Hankel, we must convert them into the closest Hankel matrices that preserve the decomposition. The step of SSA in which this is done is referred to as reconstruction, resulting in the \textit{coefficient reconstructions}.

The final step is the reconstruction of the original series from the PC groups which is referred to as the reconstruction stage. In order for a trajectory matrix to unambiguously correspond to a time series it must be Hankel. However, in general $\boldmatrix{\tilde{T}}^k $ need not be Hankel. It is thus necessary to ``Hankelise" each $\boldmatrix{\tilde{T}^k}$ by setting each anti-diagonal to the average value along the anti-diagonal.
The ``Hankelised" matrix is the best Hankel representation of the original matrix in terms of the Frobenius norm \citep[see][]{Weinberg...mSSA...2021}. This antidiagonal averaging procedure results in {\it reconstructed coefficients} for a set of pre-selected PCs, $I_m$:
\begin{eqnarray}\small
  \tilde{s}^k_j &=&
  \begin{cases} \displaystyle
    \frac{1}{j} \sum_{n=1}^{j} U^k_{n-j+1} V^k_n & \mbox{if}\ 1\le j < L-1, \\
    \displaystyle
    \frac{1}{L} \sum_{n=1}^{L} U^k_{n-j+1} V^k_n & \mbox{if}\ L\le j \le N - L + 1 \, \\
    \displaystyle
    \frac{1}{N-j+1} \sum_{n=N-L+1}^{N} U^k_{n-j+1} V^k_l & \mbox{if}\ N-L+2\le j \le N. \\
 \end{cases}\nonumber \\
 &=&\{\tilde{s}_{1,k} \dots \tilde{s}_{N,k}\}.
 \label{eq:antidiagonalaverage}
\end{eqnarray}
The series $\tilde{s}_k$, where $k$ is the group label (typically an integer), is the {\it reconstructed coefficient} series. 

\subsubsection{Power Spectra from Discrete Fourier Transforms}
\label{subsubsec:powerspectra}

We use the Discrete Fourier Transform (DFT) to estimate primary frequencies in reconstructed coefficients for the purposes of grouping. 
The DFT of a reconstructed coefficient is defined as
\begin{equation}
    \tilde{s}^k(\omega) = \mathcal{F}[\tilde{s}^k(t)] =
    \sum_{j=0}^{N-1} e^{-i2\pi \omega t_j / N} \tilde{s}^k(t_j)
\label{eq:DFT}
\end{equation}
where $C(t_k)$ is a reconstructed coefficient time series of interest. 
One may also compute the DFT of a PC directly for an alternate grouping strategy.
Plotting the resulting frequency and DFT values results in the power spectrum.

\subsubsection{Contrast measurement}
\label{subsubsec:contrast}

Using the reconstructed coefficients, one can construct the field of interest using a reduced version of either equation (\ref{eq:sphericalcoefficientprojection}) or equation (\ref{eq:fourierlaguerrereconstruction}), depending on the component.
The reduced version of each equation will sum over a selected set of indices corresponding to the input series to mSSA.
For example, if one were to use disc coefficients $C_{m=0, n\in[0,5]}$ as the input series $\vec{s}$ for the group indexed by $k$ (cf. equation~\ref{eq:antidiagonalaverage}), the surface density representation given by equation~\ref{eq:fourierlaguerrereconstruction} would become
\begin{equation}
\tilde{\Sigma}^k_{\rm disc}(R, \phi;t) = \sum_{m=0}\sum_{n\in[0,5]} \tilde{s}^k_{mn}(t)e^{im\phi}G_n(R),
\label{eq:reducedfourierlaguerrereconstruction}
\end{equation}
where the notation $\tilde{\cdot}$ indicates an approximated and truncated field representation (in this case surface density) from a select PC group.

To compute the contrast $\Delta_\Sigma$, we divide the selected field representation by the unperturbed $m=0,n=0$ representation
\begin{equation}
\hat{\Sigma}_{{\rm disc},00}(R;t) = C_{00}(t)G_0(R),
\end{equation}
giving
\begin{equation}
\Delta_\Sigma(R, \phi;t) = \frac{\tilde{\Sigma}^k_{\rm disc}}{\hat{\Sigma}_{{\rm disc},00}}.
\label{eq:contrastequation}
\end{equation}
Analogous expressions exist for computing halo density contrast. 
In this work we consider only disc surface density contrasts.

\subsubsection{Implementation of mSSA}
\label{sec:mssa_impl}

mSSA involves the exact same procedure as SSA, but with a \emph{grand}-trajectory matrix \citep{Ghil...review...2002} that is the concatenation of the trajectory matrices for the individual series. 
All other procedures are the same except for diagonal averaging. 
The choice of $L$ is no longer symmetric around $\frac{N}{2}$ because only one of the dimensions is correlated.

We form the grand trajectory matrix by concatenating trajectory
matrices:
\begin{equation}
  \boldmatrix{H} =
  \begin{bmatrix}
    \boldmatrix{T}_0,\boldmatrix{T}_1, \dots,\boldmatrix{T}_M
  \end{bmatrix}
  \in \mathbb{R}^{L\times MK}
\end{equation}\label{traj}
where $M$ is the total number of different coefficient series. The next step is to find the SVD of $\boldmatrix{H}$ such that we may write the grand-trajectory matrix as the sum of principal components, $\boldmatrix{H}=\sum_{i=1}^r \sigma_iU_iV_i^\intercal$. This could be computed directly but it is unnecessarily expensive and could be prohibitive for a large number of series. We therefore solve the smaller $L\times L$ eigenvalue problem instead.

First note that $\boldmatrix{H}\in\mathbb{R}^{L\times MK}$ and $L \leq N$, thus $\boldmatrix{H}\boldmatrix{H}^\intercal\in\mathbb{R}^{L\times L}$ is smaller than $\mathbb{R}^{N\times N}$. 
If $N$ is sufficiently modest (as is the case in this work with $N=599$), we may find $\boldmatrix{U}$ by performing the eigendecomposition
\begin{equation}
\boldmatrix{H}\boldmatrix{H}^\intercal=\boldmatrix{U}\boldmatrix{\Lambda} \boldmatrix{U}^\intercal,
\end{equation}
which only requires the eigen-decomposition of the $L\times L$ matrix $\boldmatrix{H}\boldmatrix{H}^\intercal$.
We can then find
$\boldmatrix{\Sigma}\boldmatrix{V}^\intercal$ as
\begin{equation}
\boldmatrix{U}^\intercal\boldmatrix{H}=\boldmatrix{U}^\intercal(\boldmatrix{U}\boldmatrix{\Sigma}\boldmatrix{V}^\intercal)=\boldmatrix{\Sigma}\boldmatrix{V}^\intercal.
\end{equation}
We then have everything required to write $\boldmatrix{H}=\sum_i^L U_i \sigma_i V_i^\intercal$ doing only the eigen-decomposition of an $L\times L$ matrix. 
We found that this optimisation, in the modest $N$ limit, allowed the correlation of a much greater number of input series $M$.

After obtaining $\boldmatrix{H} = \sum_i \sigma_iU_i V_i^\intercal$, we do the
same grouping described in 3.1.3 and obtain $\boldmatrix{H} = \sum_{k=1}^m
\boldmatrix{\tilde{T}}_k$ where $ \boldmatrix{\tilde{T}}_k = \sum_{i \in I_k}
\sigma_iU_i V_i^\intercal$. 
We also must Hankelize each
$\boldmatrix{\tilde{T}}_k$ 
by applying the Hankelization algorithm from equation (\ref{eq:antidiagonalaverage}) to each block of the grand trajectory matrix independently.
After this procedure has been applied, each Hankelized PC group ideally corresponds to a different feature.

\end{document}